\documentclass[nofootinbib,pre,superscriptaddress,twocolumn,longbibliography,floatfix]{revtex4-1}
\pdfoutput=1
\usepackage[utf8]{inputenc}
\usepackage{hyperref}
\usepackage{graphicx}  
\usepackage{dcolumn}   
\usepackage{bm}        
\usepackage{amssymb,amsmath}   
\usepackage{uri}
\usepackage{xr}

\usepackage[x11names, rgb]{xcolor}
\definecolor{sbase03}{HTML}{002B36}
\definecolor{sbase02}{HTML}{073642}
\definecolor{sbase01}{HTML}{586E75}
\definecolor{sbase00}{HTML}{657B83}
\definecolor{sbase0}{HTML}{839496}
\definecolor{sbase1}{HTML}{93A1A1}
\definecolor{sbase2}{HTML}{EEE8D5}
\definecolor{sbase3}{HTML}{FDF6E3}
\definecolor{syellow}{HTML}{B58900}
\definecolor{sorange}{HTML}{CB4B16}
\definecolor{sred}{HTML}{DC322F}
\definecolor{smagenta}{HTML}{D33682}
\definecolor{sviolet}{HTML}{6C71C4}
\definecolor{sblue}{HTML}{268BD2}
\definecolor{scyan}{HTML}{2AA198}
\definecolor{sgreen}{HTML}{32CD32}
\hypersetup{
  colorlinks = true,
  citecolor = {sgreen},
  urlcolor = {sblue}
}

\usepackage{footnote}
\usepackage{footmisc}

\usepackage{xcolor}
\definecolor{orange1}{rgb}{1, .467, 0}
\definecolor{blue1}{rgb}{.36, .494, .714}

\newcommand{\beginsupplement}{%
 \setcounter{table}{0}
  \renewcommand{\thetable}{S\arabic{table}}%
  \setcounter{figure}{0}
  \renewcommand{\thefigure}{S\arabic{figure}}%
  \setcounter{equation}{0}
  \renewcommand{\theequation}{S\arabic{equation}}%
}

\begin{document}
\title{Measuring amount of computation done by C.elegans using whole brain neural activity}
\author{Junang Li}
\email[Corresponding author: ]{junangl@princeton.edu}
\affiliation{Department of Physics, Princeton University, Princeton, New Jersey 08544, United States of America}
\author{Andrew M. Leifer}
\affiliation{Department of Physics, Princeton University, Princeton, New Jersey 08544, United States of America}
\affiliation{Princeton Neuroscience Institute, Princeton University, Princeton, New Jersey 08540, United States of America}
\author{David H. Wolpert}
\email[Corresponding author: ]{david.h.wolpert@gmail.com}
\affiliation{Santa Fe Institute, Santa Fe, New Mexico 87501, United States of America}
\affiliation{International Center for Theoretical Physics, Trieste I-34151, Italy}
\affiliation{Complexity Science Hub, Vienna 1080, Austria}
\affiliation{Arizona State University, Tempe, Arizona 85287, United States of America}

\begin{abstract}
Many dynamical systems found in biology, ranging from genetic circuits to the human brain to human social systems, are inherently computational. 
Although extensive research has explored their resulting functions and behaviors, the underlying computations often remain elusive. 
Even the fundamental task of quantifying the \textit{amount} of computation performed by a dynamical system remains under-investigated. 
In this study we address this challenge by introducing a novel framework to estimate the amount of computation implemented by an arbitrary physical system based on empirical time-series of its dynamics. 
This framework works by forming a statistical reconstruction of that dynamics, and then defining the amount of computation in terms of both the complexity and fidelity of this reconstruction.
We validate our framework by showing that it appropriately distinguishes the relative amount of computation across different regimes of Lorenz dynamics and various computation classes of cellular automata. 
We then apply this framework to neural activity in \textit{Caenorhabditis elegans}, as captured by calcium imaging.
By analyzing time-series neural data obtained from the fluorescent intensity of the calcium indicator GCaMP, we find that high and low amounts of computation are required, respectively, in the neural dynamics of freely moving and immobile worms. 
Our analysis further sheds light on the amount of computation performed when the system is in various locomotion states.
In sum, our study refines the definition of computational amount from time-series data and highlights neural computation in a simple organism across distinct behavioral states.
\end{abstract}

\maketitle

\section{Introduction}

From neural networks in simple organisms to modern cloud computer systems to human social systems, computational processes are pervasive in both natural and artificial systems~\cite{sourjik2012responding, navlakha2011algorithms,wolpert2024computational}. 
Centuries of research have significantly advanced our theoretical understanding of computational processes.
In particular, mathematical developments have deepened our understanding of the computability~\cite{sipser1996introduction} and complexity~\cite{arora2009computational,li2008introduction} of computational tasks, exemplified by deep issues like the famous P versus NP problem. 

Parallel to these theoretical developments, substantial progress has been made in engineering physical systems explicitly designed to implement desired computations.
Crucially, in such human engineered computers, we choose how to map the physical system's degrees of freedom to the logical variables in the abstract computer we wish to view that system as implementing. 
This means that the relationship between such a system’s dynamics and a computational process it is implementing is explicitly known before the dynamical system starts its evolution. 
In short, we have an \textit{a priori} ``computational blueprint'' for mapping the dynamics of the physical system to that of a computer.

In general though, one can identify many (often infinite) different computations with the dynamics of any given physical system~\cite{piccinini2010computation}.
As a result, naturally occurring computational systems—such as biological neural networks or genetic circuits—lack an \emph{a priori} computational blueprint. 
Instead, we are confronted by the problem of observing their computational dynamics in nature, and then inferring from those observations what computation such a system performs~\cite{urai2022large}.
In other words, we must choose one of the many different computations that are consistent with the observed dynamics, and privilege it as ``the" computation that the system is performing. 

As ill-posed as the problem of making such a choice is, it is a necessary first step to be able to analyze the computation performed by any dynamical system.
One common strategy to this problem that researchers in biology have used to grapple with this issue has been to impose specific tasks on the biological system, and then try to use its response to those tasks to infer the physical system's computational blueprint~\cite{sun2022cortical,el2024chronic,kim2017ring}.  For example, a task for a rodent might be to correctly integrate two competing stimuli in order to recieve a food reward. In that example the computation is integration. 
Note, though, that this approach implicitly assumes that the dynamics of the biological system is optimized for a goal known to the researcher, such as integrating a stimulus or maintaining homeostasis.
Yet in many cases, even the fundamental goal these systems are addressing can remain elusive, regardless of what computation they might be using to achieve such a goal. 

A hopefully more tractable version of this challenge is to quantify the \textit{amount of computation} a system performs, based on its observed dynamics. 
Rather than attempting to tackle the broad question of how to learn the precise computation, we focus instead on directly estimating the amount of computation from observed dynamics.

Here we address this challenge by developing a data-driven framework to define the amount of computation performed by dynamical systems directly from observed time-series data, without requiring explicit knowledge of the underlying task or algorithm the system implements. 
Critically, our framework does not aim to establish an absolute or objective measure of computation; instead, it assesses \textit{relative amounts of computation} across multiple systems or behavioral conditions. 
We achieve this through a Pareto-front-based analysis of the inherent trade-off between complexity and fidelity, quantified via statistical reconstruction of the observed dynamics.

To validate our framework, we use artificial systems such as Lorenz dynamics and cellular automata (CA), where consensus exists on relative computational complexity, and subsequently extend our analysis to biological neural dynamics in \textit{Caenorhabditis elegans} (\textit{C. elegans}). 
Biological neural networks provide an ideal testing ground due to their adaptive and complex behaviors, allowing us to demonstrate the practical utility of our approach in decoding biological computations.

The following sections outline our approach and validation process in detail. 
Section~\ref{sec:algorithm} introduces the conceptual workflow of our framework and the model systems used. 
In Sections~\ref{sec:Lorenz} and~\ref{sec:CA}, we validate our approach using Lorenz and CA dynamics across various time-series statistical reconstruction algorithms. 
Section~\ref{sec:immobile} then applies the framework to compare the amount of computation of \textit{C. elegans} neural dynamics in two extreme behavioral states: mobile and immobilized. 
Section~\ref{sec:rank} further demonstrates the utility of our approach by comparing the relative amount of computation across multiple behavioral conditions in \textit{C.elegans}.
Finally, in Section~\ref{sec:discussion}, we discuss broader implications, limitations of our current approach, and potential directions for future research.

\begin{figure}[htb!]
  \centering
  \includegraphics[width=1\linewidth]{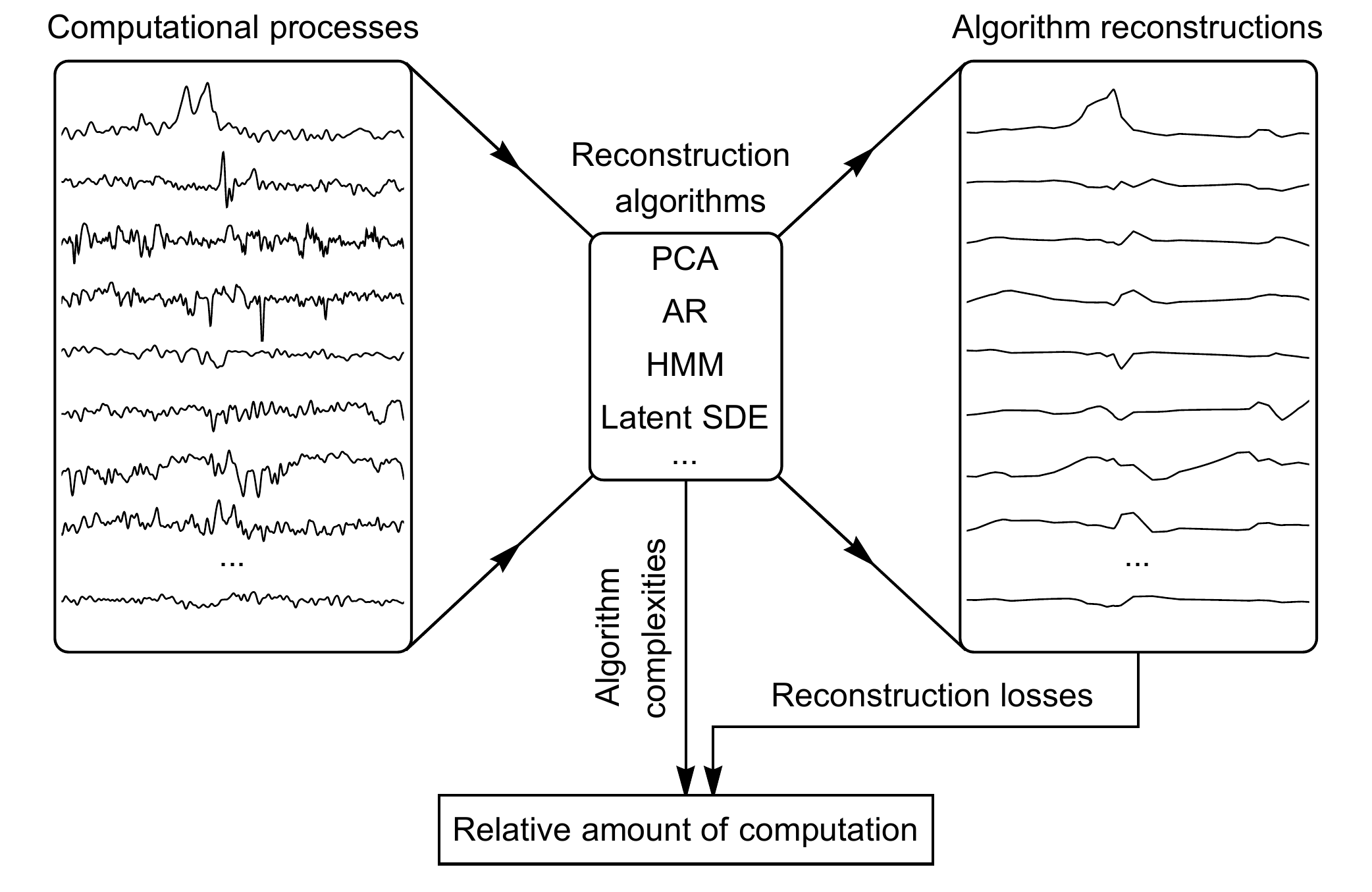}
  \caption{\textbf{Conceptual workflow for measuring the amount of computation.}
    The workflow begins with observed time series', such as time-histories of individual neuron activities.
    These are then reconstructed using a reconstruction algorithm. 
    The algorithm's complexity and the reconstruction loss collectively characterize the relative amount of computation.
    }
  \label{fig:workflow}
\end{figure}

\begin{figure*}[htb!]
  \centering
  \includegraphics[width=1\linewidth]{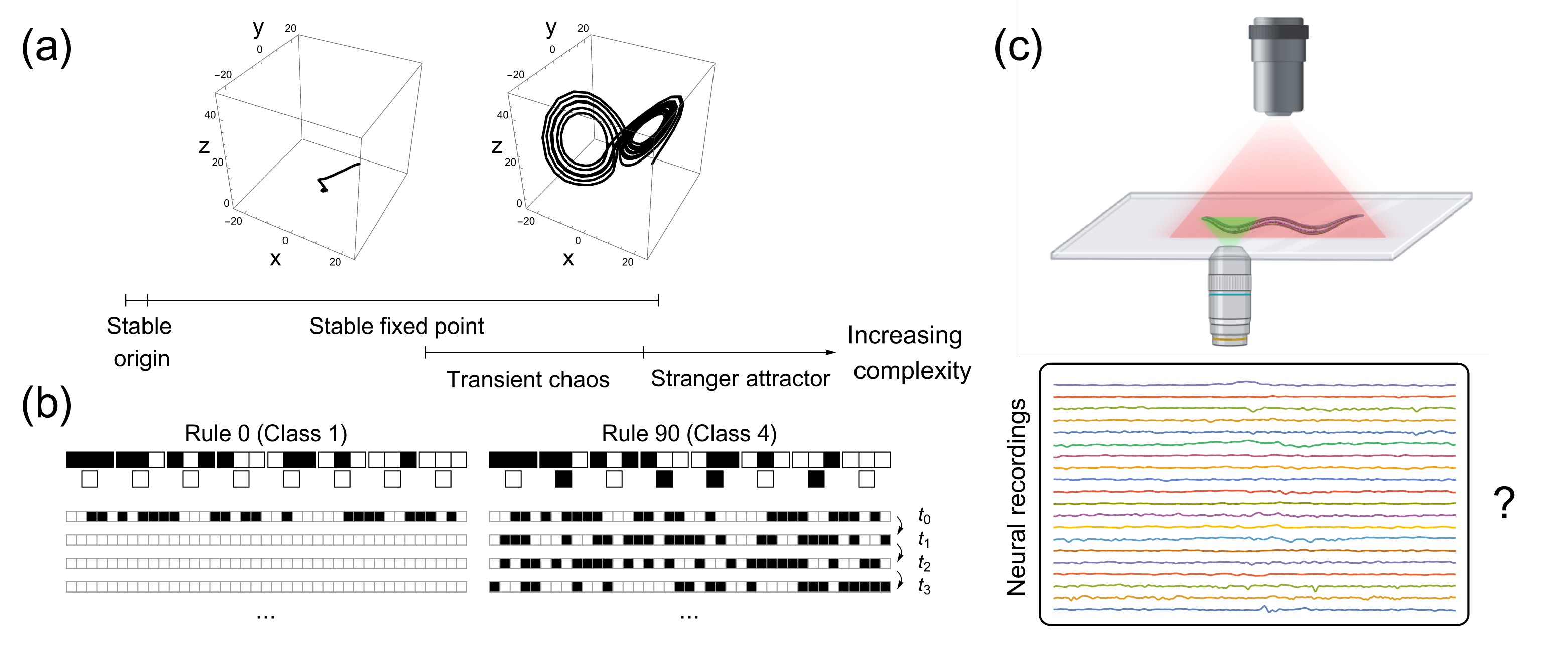}
  \caption{\textbf{Schematics of model systems for validating the computation measurement.} 
    (a) Exemplary Lorenz dynamics illustrating converging to a stable fixed point (left) and strange attractor (right). Lorenz dynamics exhibit a range of behaviors, with increasing complexity, from stable origins to limit cycles, and finally to strange attractors.
    (b) Two examples of CA rules with increasing computational complexity. The top panels illustrate all possible outcomes for a two-state nearest-neighbor CA, defining the CA rules.Given the same initial condition at $t_0$, the CA updates recursively over time, as shown in the bottom panels. Rule 0 quickly evolves into a homogeneous state (Class 1 behavior), while Rule 90 exhibits complex dynamics (Class 4 behavior).
    (c) Schematic of the experimental setup for simultaneous population recordings of \textit{C. elegans} neural activity with behavioral tracking.
    While there is a consensus on the relative amount of computation across different Lorenz dynamics (a) and CA classes (b), \textit{C. elegans} neural activity (c) remains largely unexplored.
    }
  \label{fig:model_schematics}
\end{figure*}

\section{Quantifying the amount of computation with reconstruction algorithms}\label{sec:algorithm}

Fig.~\ref{fig:workflow} demonstrates the conceptual workflow of our framework for quantifying the amount of computation in an arbitrary dynamical system, based on time-series samples of their dynamics. 
Most computational systems consist of multiple computation units, such as neurons or digital gates, that coordinate to perform computation. 
The dynamics of such systems can often be captured by relatively simple statistical models that reconstruct key features of the observed dynamics.
Moreover, capturing the dynamics of systems that perform more computation often requires increasingly complex reconstructions, reflecting the intuitive expectation that more computation demands more detailed representations of their underlying behavior.
Consequently, we hypothesize that the complexity of these statistical reconstructions of algorithms provides information about the corresponding amount of computation. (Alternatively, one might view such complexity as a \textit{definition} of the amount of computation done by an arbitrary physical system.)

Exactly capturing the complete dynamics of a system using any finite statistical reconstruction is generally impossible. 
Therefore, we need to refine our definition of the amount of computation to account for both the complexity of these statistical reconstructions and the reconstruction accuracy. 
However, this definition inherently depends on the choice of reconstruction algorithm, introducing ambiguity into our measure of complexity.
Consequently, we do not claim that our framework provides an absolute or objective measure of computation; rather, we explicitly acknowledge and systematically explore this variability in the following sections.

Importantly, in our framework, the amount of computation performed by a dynamical system is not represented by a single number, but by a Pareto front capturing the trade-off between reconstruction accuracy and complexity. 
In practice, we find that consistent comparisons of the relative amount of computation across multiple dynamical systems can be made by holding one dimension of the Pareto front fixed.
However, Pareto fronts of different dynamical systems may intersect, and these intersections can offer valuable insights into the underlying computational mechanisms—for instance, by revealing the intrinsic dimensionality of the computational dynamics, as further explored in Section~\ref{sec:Lorenz}, Fig.~\ref{fig:lorenz} and~\ref{fig:lorenz_vary}.

Since many real-world computational processes are transient and experimentally challenging to access, we focused on algorithms suitable for short time-series data. 
This requirement naturally excluded more data-intensive methods, such as delay-space embedding and nonlinear PCA, which require extensive statistical sampling.
Within this practical constraint, we selected four representative statistical reconstruction methods spanning distinct methodological axes: Principal Component Analysis (PCA)~\cite{jolliffe2016principal}, a linear method that disregards temporal dynamics; Variational Autoencoders (VAE)~\cite{kingma2013auto}, a neural network-based method designed for high-quality reconstruction but still ignoring temporal correlations; Vector Autoregression (VAR)~\cite{box2015time}, a linear method capturing first-order temporal correlations; and Latent Stochastic Differential Equations (Latent SDE)~\cite{li2020scalable,kidger2021neuralsde}, a neural network-based method explicitly modeling nonlinear temporal dynamics.

While each selected algorithm carries specific assumptions and limitations, together they enable meaningful exploration into the quantification of computation under practical constraints. 
A comprehensive evaluation of all available reconstruction methods lies beyond the scope of this study but represents an important avenue for future research.

To validate our framework for measuring the amount of computation based on reconstruction complexity and accuracy, we applied it to model systems with well-understood computational properties. 
One such system is the Lorenz system, a canonical dynamical system whose behavior varies from simple stable fixed points to chaotic strange attractors depending on system parameters [Fig.~\ref{fig:model_schematics}(a)]~\cite{doedel2015global}.
The complexity of these dynamics increases as the system transitions from stable trajectories to chaotic regimes, providing a natural benchmark for quantifying the amount of computation.

In addition to the Lorenz system, which exemplifies computation performed by a continuous dynamical process, we also considered CA as discrete computational machines capable of executing algorithms designed for specific tasks~\cite{wolfram1984cellular}. 
The 256 elementary CA consist of a linear array of binary-state cells that update synchronously based on the states of their immediate neighbors and themselves. 
Each CA rule is uniquely defined by an 8-bit binary string, resulting in 256 possible rules [Fig.~\ref{fig:model_schematics}(b)]. 
Prior studies have categorized these 256 rules into four distinct computational classes with increasing complexity, offering well-defined benchmarks for validating our approach.

Finally, extending our framework to biological systems, we examined neural dynamics in \textit{C. elegans}, an organism whose relatively simple nervous system of just 302 neurons can nonetheless produce remarkably complex behaviors [Fig.~\ref{fig:model_schematics}(c)]~\cite{kaplan2020nested, ji2021neural, hallinen2021decoding}.
Moreover, we can leverage advanced calcium imaging techniques to simultaneously monitor neuronal activity, behavior, and environmental conditions in real-time~\cite{nguyen2016whole}, making \textit{C. elegans} an ideal system to uncover principles of neural computation.

To be broadly useful, our framework should yield sensible quantification of the amount of computation across all these distinct and diverse systems.

\section{Evaluating reconstruction algorithms for quantifying computation in Lorenz dynamics}\label{sec:Lorenz}

\begin{figure}[htb!]
  \centering
  \includegraphics[width=1\linewidth]{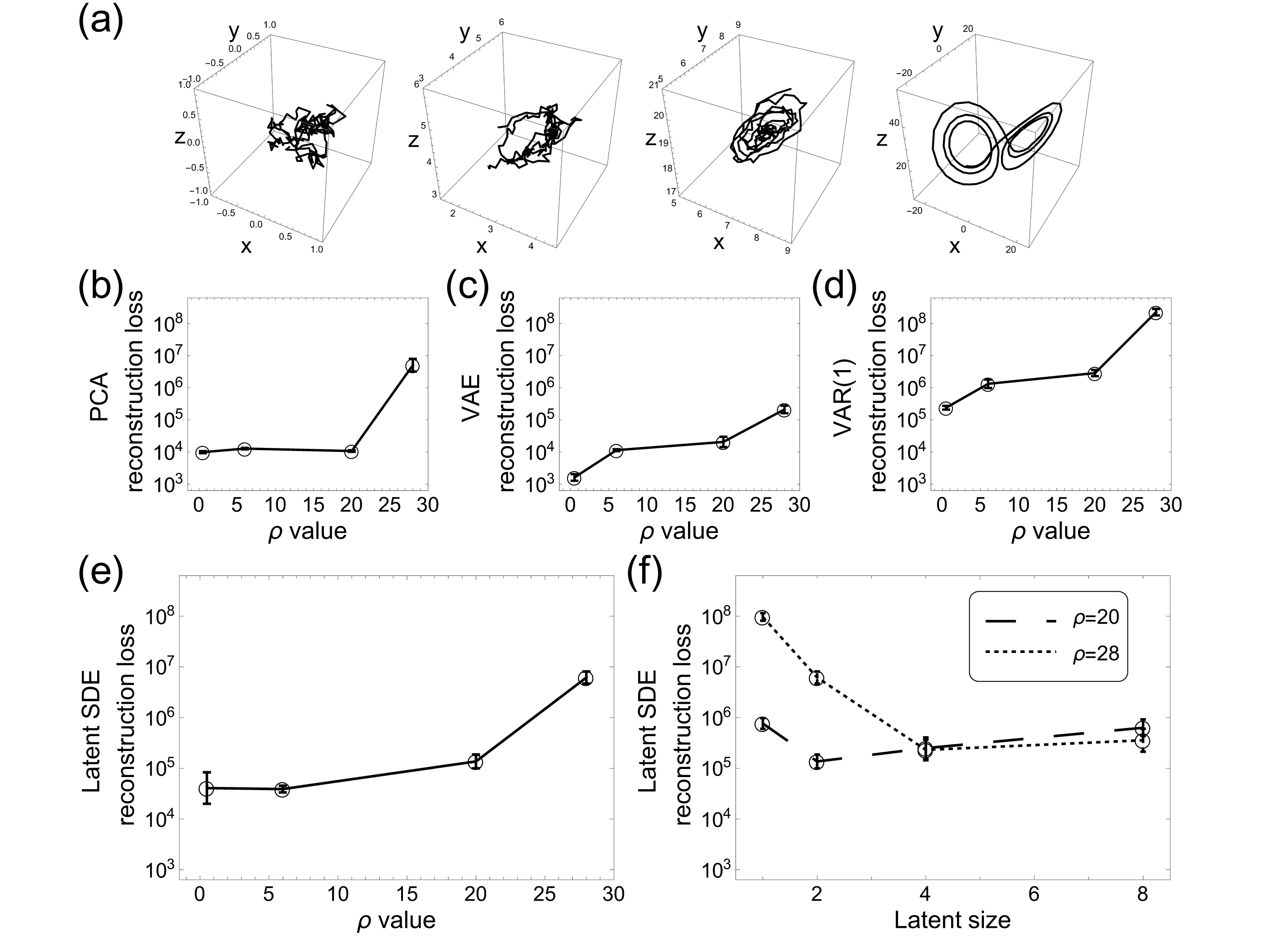}
  \caption{\textbf{Reconstruction losses capture the relative amount of computation by Lorenz dynamics.}
  (a) Examples of Lorenz dynamics across dynamical regimes with $\rho=0.5, 6, 20$, and $28$. The amount of computation performed increases from left to right.  
  (b) Reconstruction loss when using the first 2 PCs.
  (c) Reconstruction loss when using VAE with 2 latent dimensions. 
  (d) Reconstruction loss when using first-order vector AR model. 
  (e) Reconstruction loss when using Latent SDE with 2 latent dimensions. 
  Error bars represent the standard error of average based on 10 independent simulations for each Lorenz parameter.
    }
  \label{fig:lorenz}
\end{figure}

Following convention~\cite{doedel2015global}, we fixed $a=10$ and $b=8/3$ in the Lorenz equation: 
\begin{equation}
    \begin{aligned}
        \dot{x}&=a(y-x) \\
        \dot{y}&=x(\rho-z)-y \\
        \dot{z}&=xy-bz,
    \end{aligned}
\end{equation}
while varying $\rho$ to drive the Lorenz system through different dynamical regimes. 
Fig.~\ref{fig:lorenz}(a) illustrates exemplary behaviors for stochastic Lorenz simulations, transitioning from the stable regime ($\rho = 0.5$), to limit cycles ($\rho = 6$ and $\rho=20$), and finally to strange attractor ($\rho = 28$) [Methods].

Given its widespread use in neuroscience, we first analyzed these regimes using PCA. 
Since the Lorenz system is intrinsically three-dimensional, reconstructing with all three principal components (PCs) would trivially result in perfect reconstruction. 
Therefore, we focus our analysis on reconstructions using only two PCs, leaving the full analysis to the SI [Fig.~\ref{fig:lorenz_vary}].

Fig.~\ref{fig:lorenz}(b) shows PCA reconstruction loss (negative log-likelihood) across the four dynamical regimes when using two PCs. 
Reconstruction losses were indistinguishable between the stable and limit-cycle regimes, whereas the chaotic regime exhibited a significantly higher loss. 
This similarity occurs because both stable and limit-cycle Lorenz dynamics are intrinsically two-dimensional or lower, making two PCs equally sufficient for capturing their variability, irrespective of their specific dynamical behavior. 
However, by reducing the number of PCs to one, PCA reconstruction clearly distinguishes these regimes, recovering the expected trend in computational complexity [Fig.~\ref{fig:lorenz_vary} and SI].

For a nonlinear comparison, we also implemented a VAE, trained independently on individual time points of the Lorenz dynamics without explicitly modeling temporal correlations. 
Fig.~\ref{fig:lorenz}(c) shows VAE reconstruction losses, which qualitatively follow the expected trend of increasing computational complexity across the regimes. 
Notably, the VAE’s nonlinear architecture provided superior reconstruction performance compared to the other methods tested.

Next, we applied a first-order VAR model [VAR(1)], a standard linear method for modeling temporal dependencies in time-series data, to the four Lorenz dynamical regimes [Methods].
As anticipated, stable fixed point and limit cycle behaviors can be captured through local linearization and thus modeled by a linear AR approach, whereas chaotic behavior necessitates higher-order models. 
Overall, the VAR(1) correctly captures the coarse trend of the amount of computation across the regimes, with significantly increased loss for the chaotic dynamics.

Finally, Fig.~\ref{fig:lorenz}(e) shows reconstruction losses using the Latent SDE model with two latent dimensions. 
Unlike PCA and VAE, which treat dynamics as static snapshots, Latent SDE reconstructs high-dimensional dynamics as evolving trajectories within a lower-dimensional latent space governed by SDEs. 
Latent SDE reconstruction loss also consistently captures the expected ordering, correctly ranking Lorenz dynamics from stable through chaotic regimes.

Fig.~\ref{fig:lorenz}(f) further examines how reconstruction accuracy varies with latent dimension size, comparing two exemplary dynamical regimes: the limit-cycle ($\rho=20$) and chaotic ($\rho=28$). 
Initially, reconstruction loss significantly decreases with increasing latent size. 
However, after this initial decreasing, the loss for the limit-cycle regime ($\rho=20$) increases with larger latent dimensions, eventually surpassing the chaotic regime's loss at a latent dimension of 8. 
This intersection occurs because the intrinsic dimensionality differs between regimes: the limit-cycle regime is intrinsically two-dimensional, so further increasing latent dimensions introduces unnecessary complexity, causing poorer generalization and increased loss due to overfitting~\cite{goodfellow2016deep}. 
In contrast, the chaotic regime, intrinsically three-dimensional, continues benefiting from larger latent spaces. 
Thus, intersections of reconstruction losses across latent dimensions provide practical criteria for estimating the intrinsic dimensionality of the underlying dynamics.

Importantly, statistical reconstructions can never be perfect in the presence of noise. 
While sufficiently strong noise can obscure underlying computational dynamics entirely, moderate levels of noise, as demonstrated in Fig.~\ref{fig:lorenz_noise}, elevate absolute reconstruction losses but preserve the relative ordering of computational complexity across regimes.

\section{Evaluating reconstruction algorithms for quantifying computation in Cellular Automata}\label{sec:CA}
\begin{figure}[htb!]
  \centering
  \includegraphics[width=1\linewidth]{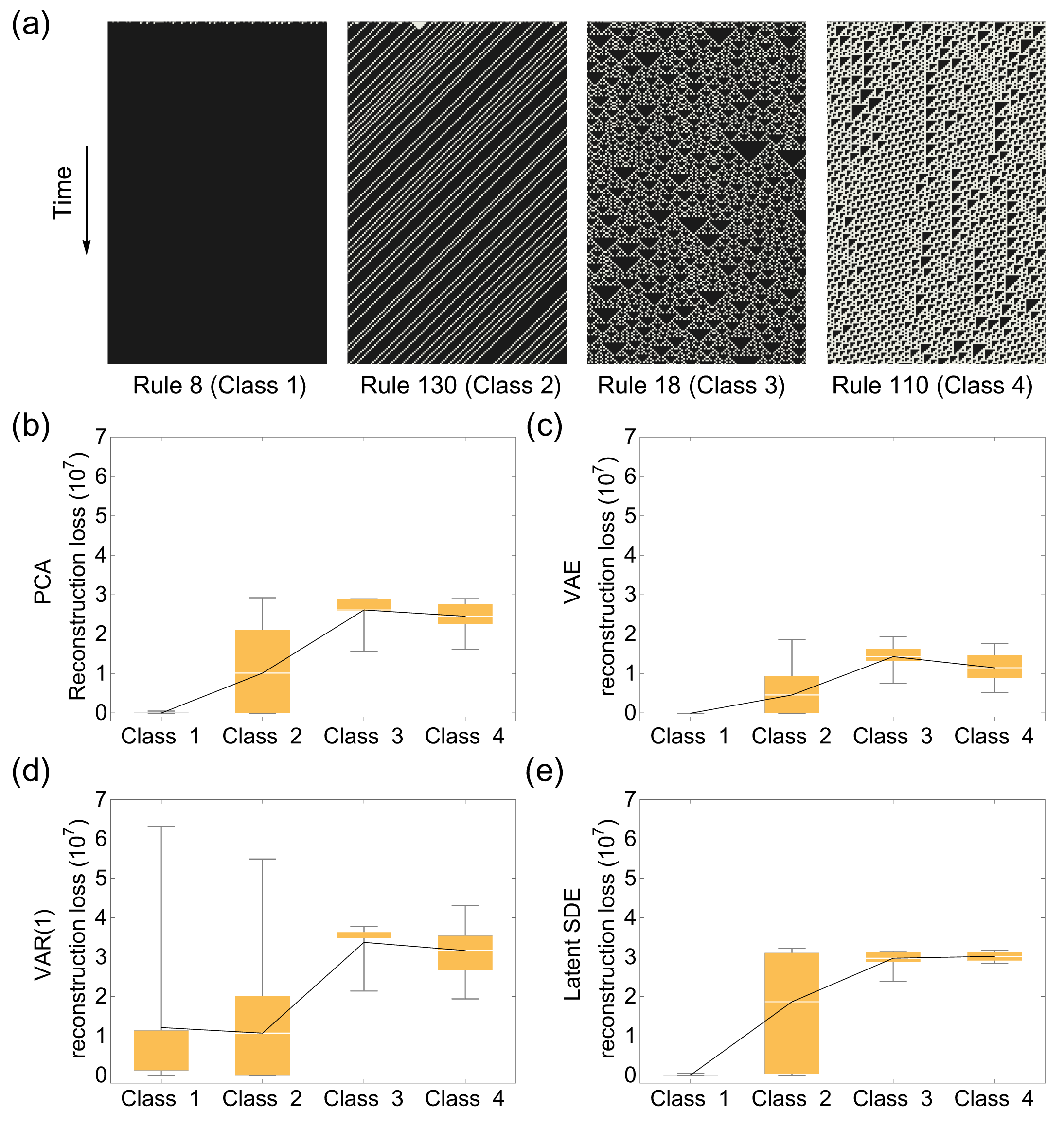}
  \caption{\textbf{Latent SDE correctly ranks the CA computation classes}. 
  (a) Examples CA dynamics for the four computation classes, with time propagating downward.
  (b) Reconstruction loss when using the first 4 PCs.
  (c) Reconstruction loss when using VAE with 4 latent dimensions.
  (d) Reconstruction loss when using first-order vector AR model. 
  (e) Reconstruction loss when using Latent SDE with 4 latent dimensions. 
  Error bars indicate the maximum and minimum values, with the yellow bar representing the 25th to 75th percentile range. The mean is shown by the white line. Results represent all CA rules within the corresponding computation class, with each rule simulated across 10 independent random initial conditions.
    }
  \label{fig:CA}
\end{figure}

To further test our measurement, we transition from continuous dynamical systems to the abstract computational models of the 256 elementary CA rules. 
Fig.~\ref{fig:CA}(a) showcases examples of space-time diagrams representing the four distinct classes of computation. 
In each diagram, the row of pixels illustrates the states of the automaton's cells at a specific time instance, with time flowing downward. 
We initialize our CAs with random arrays of 0s and 1s and propagate them for 200 time steps. 
To minimize boundary effects, all CAs are simulated with a width of 1000 cells, and only the middle 128 cells are selected for analysis.
We repeat the process with different initial conditions and average the results across multiple runs to reduce variability and ensure statistical robustness [Methods].

As expected, Class 1 rules quickly converge into a spatially homogeneous state, resulting in uniform patterns. 
Class 2 rules generate sequences of stable or periodic structures, leading to repeating patterns over time. 
Moving to Class 3, patterns become more random-looking, displaying chaotic aperiodic behavior.
Lastly, Class 4 exhibits behavior characterized by localized structures that interact in complex ways, neither entirely random nor entirely repetitive.
The computation capability increases across these classes, with Class 4 being computationally universal~\cite{cook2004universality}.

Fig.~\ref{fig:CA}(b) presents the reconstruction losses when applying PCA to CA dynamics while retaining only the first four PCs (the trend persists with differing number of PCs, as shown in the Fig.~\ref{fig:CA_vary}). 
While PCA successfully distinguishes between Class 1 and Class 2 types of computation, it fails to differentiate between Class 3 and Class 4. 
As previously mentioned, directly applying PCA by projecting the data onto its principal axes disregards temporal dynamics.
Consequently, the projections of Class 3 and Class 4 CAs lose their distinct temporal correlations and become indistinguishable.

VAEs trained on individual time instances of CAs also fail to differentiate between Class 3 and Class 4 computations [Fig.~\ref{fig:CA}(c)]. 
Although the VAE generally achieves better reconstruction accuracy, it lacks awareness of the underlying dynamics, treating each time point as an isolated sample. 
This limitation highlights that improved reconstruction accuracy does not necessarily imply a better quantification of the amount of computation.
Together, these two examples underscore the importance of incorporating temporal structure when analyzing computational processes.

To explicitly account for temporal dependencies, we applied a VAR(1) model. 
As shown in Fig.~\ref{fig:lorenz}(d), although the VAR(1) approach incorporates linear temporal correlations, it failed to reliably differentiate the higher computational classes.
This reflects its inherent limitation in modeling nonlinear interactions essential for capturing the complexity of Class 3 and 4 CA rules.

In contrast, Fig.~\ref{fig:CA}(e) demonstrates that Latent SDE successfully ranks all four CA computation classes. 
The neural networks used to parameterize the SDEs go beyond the linear assumptions of AR models and have access to the entire temporal evolution. 
This approach allows for a more accurate and nuanced ranking of the computational classes, effectively capturing the intricate dynamics of Class 3 and Class 4 CAs.

As we incrementally increased the latent dimension of Latent SDE model, the observed trend persisted, albeit with a small decrease in absolute loss [Fig.~\ref{fig:CA_vary}]. 
It is noteworthy that even with high latent dimensions, such as 64, certain CA rules exhibit persistently elevated reconstruction losses. 
Furthermore, some realizations of Class 2 CAs display extremely high reconstruction losses. 
This discrepancy arises from the difference between SDEs, which are continuous time models, and CAs, which involve discrete state transitions. 
The abrupt transitions or oscillations between successive time steps in CA dynamics pose challenges for continuous models like SDEs [Fig.~\ref{fig:CA_class2}]. 
Despite the difficulty of capturing the exact dynamics, achieving comparable reconstruction losses demands higher-dimensional Latent SDEs for higher computation classes, supporting our assumption that the Latent SDE dimension serves as an indicator of the amount of computation.

It is crucial to highlight that being classified as a Class 4 rule does not guarantee the emergence of Class 4 computation behavior from any conceivable initial configuration. 
A well-known example is Rule 54, which is classified as Class 4 but,  when started from a single initial point, yields an ordered pattern [insets in Fig.~\ref{fig:CA_rule54}]. 
Our method effectively captures the sensitivity of the computation to initial conditions; by gradually increasing the complexity of the initial conditions for Rule 54, we observe that it eventually exhibits the characteristic Class 4 dynamics [Fig.~\ref{fig:CA_rule54}].

In summary, our systematic analysis of CA dynamics using these representative reconstruction algorithms underscores the importance of explicitly modeling temporal correlations and nonlinearity for quantifying the amount of computation. 
Among the methods tested, the Latent SDE model alone consistently differentiated among the computational classes [Table.~\ref{tab:methods}]. 
Thus, we adopt the Latent SDE model as our primary tool for subsequent analyses of neural dynamics.

\begin{figure}[htb!]
  \centering
  \includegraphics[width=1\linewidth]{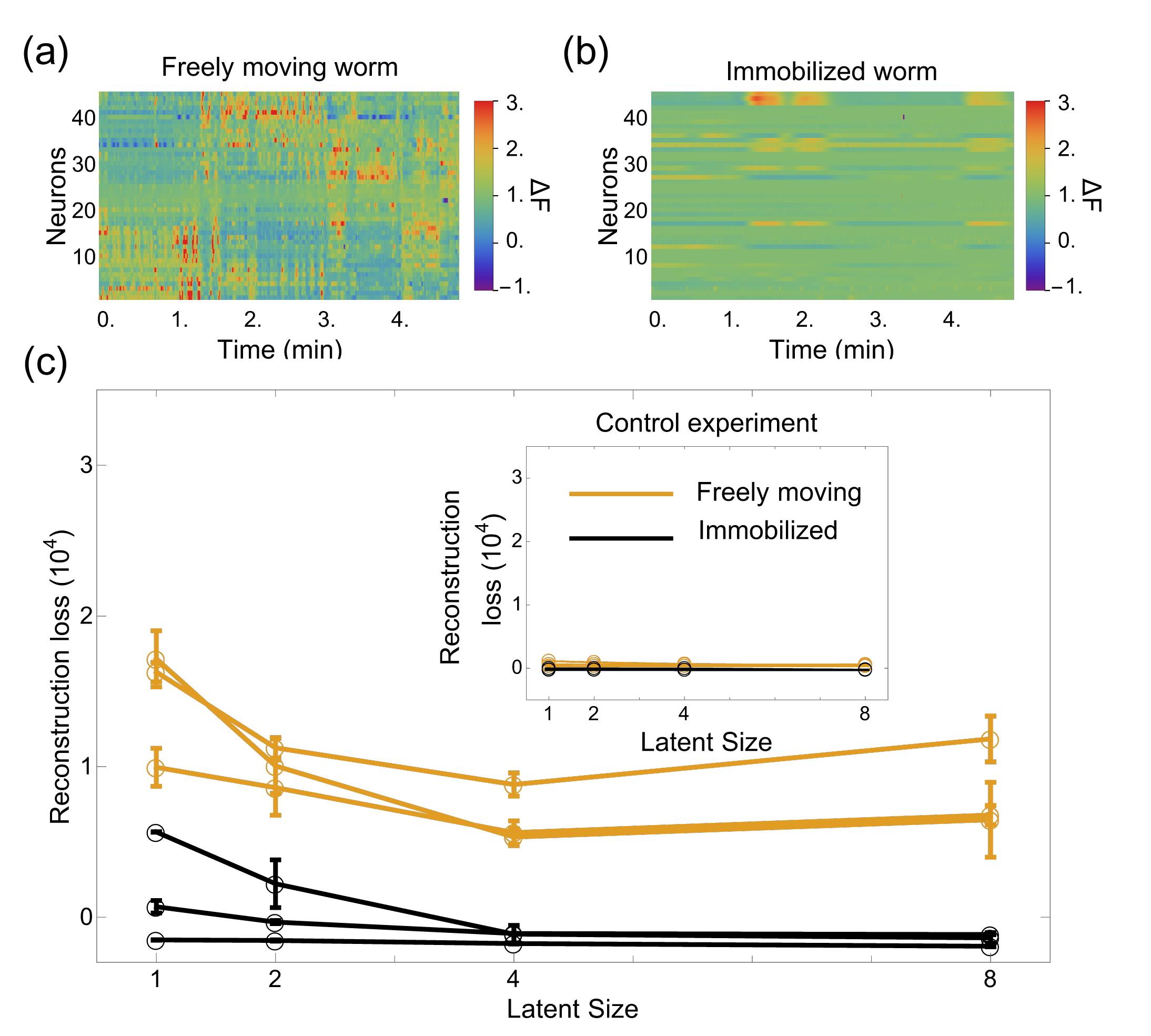}
  \caption{\textbf{Higher amount of computation measured in freely moving worms.}
  (a) Example neural recording of a freely moving worm.
  (b) Example neural recording of an immobilized worm. 
  Neurons are registered using the NeuroPAL identification system so that the same row in both mobile and immobile worms corresponds to the same neuron. 
  (c) Latent SDE reconstruction loss as a function of latent dimensions. 
  The yellow line represents the reconstruction loss for the freely moving worm's neural activity, while the black line represents that for the immobilized worm. 
  The inset shows the same analysis for GFP control experiments.
  Each line corresponds to a different animal, and error bars indicate the standard error of the mean across three different training seeds.
    }
  \label{fig:worm_state}
\end{figure}

\begin{figure*}[htb!]
  \centering
  \includegraphics[width=1\linewidth]{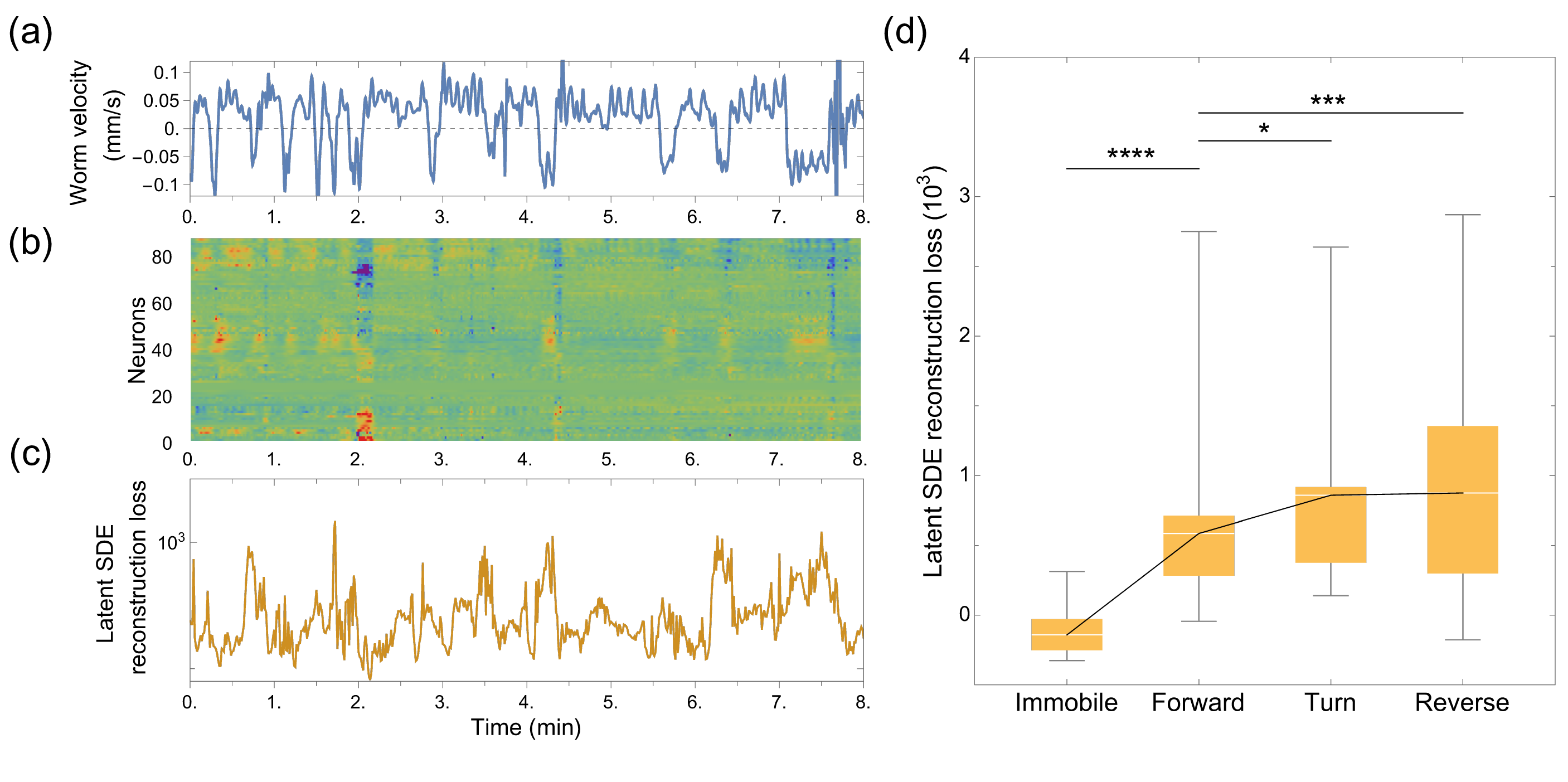}
  \caption{\textbf{Latent SDE loss rank the relative amount of computation for different worm locomotion states.}
  (a) Worm velocity projected along the body axis, indicating forward and reversal motions.
  (b) The corresponding neural activity during worm locomotion.
  (c) Latent SDE reconstruction loss over time.
  (d) Latent SDE reconstruction loss for different locomotion behaviors using 4 latent dimensions. 
  Error bars indicate the maximum and minimum values, with the yellow bar representing the 25th to 75th percentile range. The mean is shown by the white line.
  Statistical significance of the differences between behavioral states was assessed using t-tests, with significance levels indicated as follows: * for $p < 0.05$, ** for $p < 0.005$, *** for $p < 0.0005$, and **** for $p < 0.00005$.
    }
  \label{fig:worm_behavior}
\end{figure*}

\section{Detecting different amounts of computation from mobile and immobilized C.elegans}\label{sec:immobile}

The dynamic patterns of brain activity in an animal are expected to encode crucial information about its behavior. 
Despite having only 302 neurons, \textit{C.elegans} exhibits a wide repertoire of stereotypical behavioral states.
Among these states, the two extreme examples are its natural, freely moving state and an immobilized state induced by a paralytic drug.
Previous studies have highlighted notable distinctions in the neural activity corresponding to these two states~\cite{hallinen2021decoding, kato2015global, gauthey2024light}. 
These distinct states provide a unique opportunity to investigate the underlying neural computations and dynamics under different physiological conditions.

In our experiments, we employed calcium imaging to monitor the activity of most of the neurons in the head region of \textit{C.elegans} [Methods]. 
Although our recordings do not encompass the entire nerves system, they capture most interneurons, which are crucial for information processing~\cite{sabrin2020hourglass}. 
Additionally, we implemented NeuroPAL~\cite{yemini2021neuropal} to identify individual neurons and facilitate cross-animal comparisons [Method, Fig.~\ref{fig:NeuroPAL}]. 
Finally, a low-magnification imaging system is designed to track the worm motion and capture its posture [Fig.~\ref{fig:model_schematics}(c), Methods].

Fig.~\ref{fig:worm_state}(a) and (b) display example neural recordings from a freely moving worm and an immobilized worm, respectively.
Each row represents the activity of a specific neuron over the recording period.
Visually, the neural activity patterns appear distinct, with neurons in the freely moving worm displaying more frequent activations, while the immobilized worm exhibits slower and sparser activations. 
Consequently, a greater dimension is required to fit a Latent SDE model for the neural dynamics of the freely moving worms compared to the immobilized worms [Fig.~\ref{fig:worm_state}(c)]. 
This observation aligns with expectations, as a freely moving worm is likely to engage in more computation due to constant changes in its surrounding environment and exhibits various behaviors.

Importantly, motion-induced artifacts and measurement noise can introduce a `ghost' amount of computation that does not originate from neural dynamics. 
To mitigate this, we implemented a motion-correction algorithm to minimize motion-induced artifacts in the calcium imaging data [Methods, Fig.~\ref{fig:ethogram_tmac}]. 
Additionally, we performed control experiments using an animal that expressed a calcium insensitive GFP instead of calcium indicator.
These control experiments are designed to contain only motion-induced artifacts and no signal. 
They showed significantly reduced reconstruction losses for both freely moving and immobilized worms, with little decrease upon increasing the latent dimension [insert Fig.~\ref{fig:worm_state}(c)]. 
This finding confirms that our observed differences in computation are not artifacts of the imaging process but reflect genuine neural activity.

\section{Comparing the amount of computation for C.elegans behavioral states}\label{sec:rank}

Beyond its two extreme states, \textit{C.elegans} exhibits a rich spectrum of locomotion behaviors. 
One of the most salient is the animal's velocity, its direction and speed of movement [Methods]. 
Fig.~\ref{fig:worm_behavior}(a) illustrates a typical velocity profile of a freely moving worm. 
For the majority of the time, the velocity is positive, indicating forward motion, interspersed with brief intervals of negative velocity corresponding to reversal movements. 
This shift from forward to reversal behavior is also reflected in the neural activity [Fig.~\ref{fig:worm_behavior}(b)], where distinct sets of neurons are activated during reversals. 
Previous studies have established that \textit{C.elegans} engages different neural circuits and performs distinct computations during forward and reversal motions~\cite{atanas2023brain, hallinen2021decoding}; however, the relative computational complexity of these behaviors remains unclear.

To quantify the computation associated with these different behaviors, we segmented the neural recordings into overlapping 15-second subsections. 
Rather than training a single model and then predicting on hold-out data, which could bias the predictions toward more frequent behavioral states, we fitted independent, structurally identical Latent SDE models to each segment and estimate the relative amount of computation associated with the behavior occurring within the same time window. 
Fig.~\ref{fig:worm_behavior}(c) presents the Latent SDE reconstruction losses across the entire recording for various latent dimensions. 
Consistently, we observed higher reconstruction losses during reversal phases compared to forward phases, suggesting that the neural computation involved in reversal is more complex and demanding than that during forward motion. 
Based on this finding, we hypothesize that forward motion, being the worm's default state, may require less computation, while reversal, a rarer and more deliberate action, likely requires more complex processing. 

Further refinement of the behavioral states can be achieved by examining the temporal dynamics in the worm's posture space. 
For instance, negative velocity might indicate a direct reversal, where the worm simply retraces its path by reversing its bending pattern, or a turn, which involves a sharp reversal followed by a change in the direction of forward movement. 
Building on previous research~\cite{nguyen2016whole}, we categorized the freely moving worm's behavior into three distinct states: forward, reverse, and turn, based on its velocity and body posture [Methods]. 
We then applied our framework to rank the amount of neural computation involved in each of these states, along with the immobilized state [Fig.~\ref{fig:worm_behavior}(d)].  
The highest amount of computation was observed during reversal and turn, followed by forward motion, with the immobilized state exhibiting the least computation .
This ranking is robust across different lengths of neural recording segments Figs.~\ref{fig:ethogram_datalen} and cannot be trivially attributed to noise or motion artifact Figs.~\ref{fig:ethogram_tmac} and~\ref{fig:ethogram_GFP}.
The ability of our framework to discern these subtle differences in neural computation across various behavioral states underscores its potential as a powerful tool for investigating the neural basis of behavior in \textit{C.elegans} and beyond.

\section{Discussion}\label{sec:discussion}

In this study, we introduced a novel metric to quantify the relative amount of computation within arbitrary computational processes by considering both the accuracy and complexity of statistical reconstructions. 
We validated the robustness and applicability of this metric on  various computational systems where the  ordering of the relative  amount computation is already well-established. 
Among the models tested, the Latent SDE model consistently captured the expected trends, proving particularly effective due to its ability to incorporate temporal dynamics and preserve nonlinear interactions.

Leveraging this, we applied our Latent SDE-based metric to neural activity in \textit{C. elegans}. 
Our analysis consistently revealed differing amounts of computation across different locomotion behaviors, enabling us to rank these behaviors by their computational demands. 
This ranking not only offers a novel perspective to our understanding of how neural dynamics underpin behavior but also opens a new avenue for studying neural computation in a task-independent manner. 

Traditional approaches often rely on carefully designed tasks, which necessitate artificial, human-constructed environments~\cite{o2022direct}. 
In contrast, our measurement framework only requires passive observation of the computation process, enabling the study of neural dynamics in more naturalistic settings~\cite{krakauer2017neuroscience, datta2019computational}. 
This shift away from task-specific studies reduces the risk of imposing human biases on the experimental design and allows for an arguably more relevant investigation of the underlying neural mechanisms. 
Additionally, relating computation to task complexity can sometimes be misleading, particularly in scenarios where multiple solutions exist for the same task. 
By focusing on the computation itself, rather than the task's design, our approach offers a powerful tool for uncovering hidden degeneracies in neural computation.

Another promising application of our approach lies in bridging the gap between structural and dynamical complexity in neural systems. 
The connectomes of various animals, or portions thereof, are being reconstructed with increasing speed and precision~\cite{van2013wu,Dorkenwald2024}. 
Static structural complexity, as represented by these connectomes, is believed to correlate with the dynamical complexity of neural activity, which reflects  the amount of computation an organism can perform. 
Our method provides a quantitative measurement for dynamical complexity, offering a way to empirically test the relationship between an organism's connectome and its computation capabilities. 
This approach has the potential to deepen our understanding of how structural features of the brain contribute to its functional capacity.

Finally, when considering the biological brain as a physical apparatus, neural computation is constrained by physical laws, including the second law of thermodynamics, which governs the energetic cost of computation~\cite{wolpert2023stochastic,sole2024fundamental}. 
Our study of the amount of computation offers a link from the physics of computation to its functional aspects. 
By quantifying the computational demands of neural processes, our metric may help elucidate how physical principles, such as energy efficiency, influence the evolution and operation of neural systems. 
This connection between physics and function could provide new insights into the energetic constraints that shape neural computation and, by extension, behavior.

However, we acknowledge the limitations inherent in our approach. 
As illustrated in our study, the precision of our metric is highly contingent on the choice of the statistical reconstruction algorithm and the extent of prior knowledge about the system being analyzed. 
Although the Latent SDE model emerged as the most reliable among the tested algorithms, it is not without its limitations. 
Specifically, due to its underlying structure, the Latent SDE model falls short in capturing certain discretized high-frequency dynamics, such as those exhibited in some CAs. 
Moreover, the reliance on prior knowledge underscores the necessity for developing more adaptive methodologies capable of generalizing across diverse neural architectures and computational paradigms. 
There are also many other algorithms worth considering, such as nonlinear PCA, manifold learning, and various time-series prediction methods, especially those deep learning-based approaches.
Most ambitiously, one could aim to solve for the optimal algorithm based on an underlying mathematical theory of what it means for a dynamical system to compute~\cite{wolpert2024focus}. 
These limitations highlight the difficulty in developing a universal metric for quantifying the amount of computation across heterogeneous systems.

Our work constitutes a pioneering empirical exploration into the quantification of computation. 
By introducing a novel metric and applying it to a biological system, we have established a foundational framework for future research in this area. 
As an initial step, we intend this study to serve as a catalyst for further research and discourse within the neuroscience and biophysics community. 
By acknowledging these limitations and outlining potential avenues for improvement, we hope to inspire collaborative efforts toward developing more principled and universally applicable methods for understanding neural computation.

\section{Methods}
\subsection{Numerical simulation of stochastic Lorenz dynamics}
We simulate the stochastic Lorenz dynamics using a combination of deterministic and stochastic approaches, depending on the level of noise introduced.
As described in Section~\ref{sec:Lorenz}, we fixed $a=10$ and $b=8/3$ and chose $\rho$ values from $[0.5,6,20,28]$.
For deterministic simulations ($\xi=0$), we solved the Lorenz equations using the ODE solver \texttt{solve\_ivp} from the \textit{SciPy} package.
When noise was added ($\xi>0$), we used the SDE solver \texttt{sdeint.itoint} to integrate the system. 
Noise levels were varied across $[0,0.1,0.5,5]$.
Initial conditions were randomly generated around the fixed point for the corresponding parameters. 
To minimize the effect of initial condition, simulations were conducted over 1200 time points spanning $t=[0,30]$ and only the last 200 time points were taken for analysis. 
To ensure robust statistical analysis, each condition was repeated 10 times.

\subsection{Simulation of CA dynamics}
The 256 elementary CA rules can be categorized into 4 computational classes, comprising a total of 89 distinct rules (excluding symmetric pairs). 
These include 8 rules in Class 1, 66 in Class 2, 11 in Class 3, and 4 in Class 4. 
Each rule was evolved using the \textit{cellpylib} Python library, which applies the corresponding NKS rule. 
The system’s evolution followed standard CA dynamics, with state updates based on local neighborhood interactions at each time step.
For random initialization, each cell in the initial configuration was randomly assigned a value of 0 or 1, reflecting stochastic conditions. 
In the case of localized initialization, a predefined number of cells in the center of the grid were set to 1. 
The simulations were run on a grid of 1000 spatial cells for 200 time steps, with data collected from the central 128 cells. 
Each condition (a combination of rule and random initial condition) was repeated 10 times to account for variability.

\subsection{Fitting vector AR models}
We fitted the first-order vector AR model by solving a multivariate linear regression problem. 
We first constructed a lagged data matrix $\mathbf{X}$ consisted of the system's state at the previous time steps, and a matrix $\mathbf{Y}$ contained the corresponding current state. 
\begin{equation}
    \begin{aligned}
    \mathbf{X} &= [\textbf{x}_{t-1}, \textbf{x}_{t-2}, \dots] \\
    \mathbf{Y} &= [\textbf{x}_{t}, \textbf{x}_{t-1}, \dots],
    \end{aligned}
\end{equation}
where $\textbf{x}_{t}$ represents the system's state vector at time $t$.
The coefficient matrix $A$ that maps the lagged variables $\mathbf{X}$ to the current state $\mathbf{Y}$ was estimated by minimizing the residual sum of squares:
\begin{equation}
    \mathbf{Y}=A\mathbf{X} + \epsilon,
\end{equation}
where $\epsilon$ is the error term. 
The least-squares solution was obtained using the \texttt{np.linalg.lstsq} from the \textit{NumPy} library in Python, which returns the coefficient matrix $\hat{A}$ that best fits the data.
After fitting the model, we used the estimated coefficient matrix $\hat{A}$ to reconstruct the system's dynamics. 
Starting with the initial condition for the first time step, the future states of the system were iteratively predicted using the model:
\begin{equation}
    \begin{aligned}
        \hat{\textbf{x}}_{1} &= \hat{A} \textbf{x}_{0} \\
        \hat{\textbf{x}}_{2} &= \hat{A} \hat{\textbf{x}}_{1} \\
        &\dots
    \end{aligned}
\end{equation}
To prevent divergence caused by any eigenvalues of $\hat{A}$ exceeding 1, we normalized the reconstructed data to the range $[0,1]$ before computing the reconstruction error.

\subsection{VAE Architecture}

The VAE architecture comprises an encoder, a latent representation, and a decoder. 
The encoder compresses the input data $\textbf{x}$ into a latent Gaussian distribution, producing the mean $\mu$ and log variance $\log\sigma^2$ through a series of fully connected layers with LeakyReLU activations. 
The reparameterization trick is employed to sample latent variables $\textbf{z}$ from this distribution, ensuring differentiability.

The decoder reconstructs the input data from $\textbf{z}$ using a neural network with similar architecture to the encoder. 
A normal distribution is defined over the reconstructed data, and the negative log-likelihood of the observations forms the reconstruction loss. 
The latent space is regularized by a prior Gaussian distribution, and the KL divergence between the approximate posterior 
$q(\textbf{z}|\textbf{x})$ and the prior $p(\textbf{z})$ contributes to the total loss.

The loss function is the sum of the reconstruction loss and KL divergence. 
The reconstruction term encourages fidelity to the input, while the KL divergence regularizes the latent space to prevent overfitting. 
The model is trained using the Adam optimizer with an exponentially decaying learning rate, and gradient clipping ensures stability during optimization. 

\subsection{Latent SDE architecture}

The Latent SDE model architecture is adopted from~\cite{li2020scalable}, which consists of an encoder network, latent dynamics, and an observation model. 
The encoder processes the observed data sequence $\textbf{x}_{0:T}$ using a gated recurrent unit (GRU) network that operates in reverse temporal order to extract contextual information, resulting in context vectors $\textbf{c}_{0:T}$. 
A linear layer maps the GRU outputs to the desired context size. 
We approximate the posterior distribution of the initial latent state $q(\textbf{z}_0|\textbf{c}_0)$ using the context vector at the initial time step 0, sampling $\textbf{z}_0$ from this distribution.

The latent dynamics are governed by prior and posterior drift functions. 
The prior drift function $f(t,\textbf{z}_t, \textbf{c}_t)$ depends on both the latent state and the context state, parameterized by a neural network that concatenates these inputs and processes them through layers with Softplus activation functions. 
The posterior drift function $h(t, \textbf{z}_t)$ depends solely on the latent state and has a similar neural network architecture without the context input. 
The diffusion function $g(t, \textbf{z}_t)$ is diagonal, with each element modeled by a neural network ensuring positive outputs via a Sigmoid activation function. 
Lastly, a linear projection maps the latent states $\textbf{z}_{0:T}$ to the observed data space, producing reconstructions $\hat{\textbf{{x}}}_{0:T}$.

The latent states evolve over time according to the SDEs:
\begin{equation}
    \begin{aligned}
        &\text{Prior dynamics: } d\textbf{z}_t = f(t, \textbf{z}_t, \textbf{c}_t) dt + g(t, \textbf{z}_t) d\textbf{w}_d \\
        &\text{Posterior dynamics: } d\textbf{z}_t = h(t, \textbf{z}_t) dt + g(t, \textbf{z}_t) d\textbf{w}_d
    \end{aligned}
\end{equation}
where $d\textbf{w}_t$ represents a Wiener process. 

The loss function comprises two components: the negative log-likelihood of the observations given the reconstructions, and the pathwise KL divergence between the posterior and prior dynamics.
\begin{equation}
    \begin{aligned}
        &-\sum_{t=0}^{T} \ln \ p_{\hat{\textbf{x}}_{t}}(\textbf{x}_{t}) \\
        &+\lambda*\mathbb{E}_{\textbf{z}_{0:T}}\{ KL[p_{\rm prior}(\textbf{z}_{0:T}) || p_{\rm post}(\textbf{z}_{0:T})] \}
    \end{aligned}
    \label{eq:objective}
\end{equation}
The parameter $\lambda$ acts as a control parameter and is determined by a linear scheduler, annealing over epochs.
We minimize the total loss (negative ELBO) using the Adam optimizer with gradient clipping to prevent exploding gradients. 

\subsection{C. elegans preparation and whole brain imaging}
C. elegans were cultured at $20^\circ$C on nematode growth media (NGM) plates seeded with \textit{E. coli} OP50. 

Whole-brain imaging of freely moving animals was performed as described previously~\cite{nguyen2016whole}, with modifications.
A single young adult worm was transferred to a custom imaging plate composed of modified NGM media (lacking cholesterol and containing agarose instead of agar) overlaid with 10 $\mu$L of mineral oil. 
A coverslip was placed on top and mounted to the plate with valap. 
Imaging was carried out using a custom-built whole-brain imaging system that simultaneously captured the worm’s behavior, neuronal calcium activity, and panneuronal fluorescence signals.

Body posture was recorded using a low-magnification 10$\times$ brightfield objective with infrared illumination, acquiring images at 25 frames per second. 
A pre-trained SLEAP-based~\cite{pereira2022sleap} posture detection algorithm was employed in real time to track the worm’s brain position. 
A motorized stage utilized the tracking data to compensate for brain motion relative to the imaging field of view in a closed-loop system, enabling continuous tracking during movement.
High-resolution neuronal activity in the worm’s head was imaged using two 40$\times$ magnification fluorescence image streams: one for the panneuronal marker (tagRFP or mNeptune, excited at 561 nm) and the other for the calcium indicator GCaMP (excited at 505 nm). 
High-speed imaging was conducted at 200 optical slices per second, achieving a final acquisition rate of 6 head volumes per second.

For worm immobilization, we treated the worm with 10 $\mu$L of 100 $\mu$M levamisole, placed a glass slide over it, sealed the setup with valap, and proceeded with imaging using the same steps as described above.

\subsection{Multi-color imaging and neural identification}
After completing the whole brain imaging experiment, freely moving worms would be picked and immobilized for multicolor imaging. 
Volumetric, multi-color imaging was then performed to capture fluorescence signals from the NeuroPAL transgene~\cite{yemini2021neuropal}, specifically the fluorophores mTagBFP2, CyOFP1.5, tagRFP-T, and mNeptune2.5.
Channel-specific filters mounted on a mechanical filter wheel were used in conjunction with synchronized mechanical shutters to alternate laser illumination for each fluorophore. 
mTagBFP2 was imaged using a 405-nm laser with a Semrock FF01-440/40 emission filter. CyOFP1.5 was imaged using a 505-nm laser with a Semrock 609/54 emission filter. tagRFP-T was imaged using a 561-nm laser with a Semrock 609/54 emission filter, and mNeptune2.5 was imaged using a 561-nm laser with a Semrock 732/68 emission filter. 

\subsection{Neuron activity extraction}
The post-processing procedures were adapted from previously established methods~\cite{hallinen2021decoding, Nguyen2017}, with some modifications. 
Briefly, two fluorescent channels were spatially aligned using calibration beads, ensuring accurate registration across channels. 
Temporal synchronization between the high- and low-magnification imaging systems was achieved using light flashes as timing references.

High-magnification fluorescent images were processed to align the worm's body posture. 
Neural dynamics were extracted by segmenting neuronal nuclei in the red channel (RFP), and neuronal identities were assigned over time through iterative clustering. 
This method ensured consistent tracking of neurons despite motion or deformation of the worm. 
The GCaMP fluorescence signal was then extracted using the spatial positions of the neuronal nuclei obtained from tracking. 
This processing pipeline produced comprehensive datasets containing RFP and GCaMP fluorescence values for each successfully tracked neuron throughout the imaging session.

To reduce motion artifacts, the two-channel Motion Artifact Correction (TMAC)~\cite{Creamer2022} algorithm was applied. 
The final output is the artifact-corrected GCaMP time series inferred by motion reference of the RFP, presented as time-series data for individual neurons. 

\subsection{Behavioral analysis of C.elegans}
A customized U-Net model was trained to segment the worm from low-magnification brightfield images. 
The worm’s centerline was extracted by skeletonizing the segmented image. 
The head position was determined using SLEAP-labeled data, which was utilized to orient the centerline.

The worm’s velocity was quantified as the dot product between its movement direction and orientation vector. 
The center-of-mass (CoM) was calculated by combining the stage position with the CoM of the centerline. 
Positional data were smoothed and differentiated using a Gaussian kernel to compute velocity components. 
The worm’s orientation was determined from the vector connecting the head tip to a point 15\% of the centerline length away from the head tip. 
The resulting velocities were resampled to match the temporal resolution of the neural recordings.

To generate the ethogram, the worm’s CoM velocity was classified into forward, backward, and non-moving states based on the sign and magnitude of the smoothed velocity vector. 
Frames with near-zero velocity magnitudes were assigned to the non-moving state.

Non-moving states were further categorized into pausing and turning behaviors by analyzing body curvature using the third eigenworm mode. 
The projection onto the third eigenworm, which increases during deep body bends characteristic of turns~\cite{Stephens2008}, was normalized by subtracting the mean and replacing missing values with zeros. 
Turning events were identified when the projection exceeded two standard deviations from the mean or an absolute value of 10.
Short-duration behaviors ($<2.5$ s) were excluded using connected components analysis, except for reversals, which were preserved. 
Missing data resulting from excluded behaviors were interpolated using nearest-neighbor methods to ensure continuity. 
The final ethogram classified the worm’s behavior over time into forward, backward, pausing, or turning states.

\begin{acknowledgments}
D.H.W thanks the Santa Fe Institute for support.
J.L. was supported by National Science Foundation through Center for the Physics of Biological Function Grant No. PHY-1734030.
Research reported in this work was supported by the National Science Foundation, through an NSF CAREER Award to A.M.L (IOS-1845137)
\end{acknowledgments}

\section*{Author Contributions}
J.L., A.M.L, and D.H.W designed research. J.L. performed research. J.L., A.M.L, and D.H.W. wrote the paper.

\bibliography{main}

\newcommand{\noopsort}[1]{} \newcommand{\printfirst}[2]{#1} \newcommand{\singleletter}[1]{#1} \newcommand{\switchargs}[2]{#2#1}
\begin{thebibliography}{41}%
\makeatletter
\providecommand \@ifxundefined [1]{%
 \@ifx{#1\undefined}
}%
\providecommand \@ifnum [1]{%
 \ifnum #1\expandafter \@firstoftwo
 \else \expandafter \@secondoftwo
 \fi
}%
\providecommand \@ifx [1]{%
 \ifx #1\expandafter \@firstoftwo
 \else \expandafter \@secondoftwo
 \fi
}%
\providecommand \natexlab [1]{#1}%
\providecommand \enquote  [1]{``#1''}%
\providecommand \bibnamefont  [1]{#1}%
\providecommand \bibfnamefont [1]{#1}%
\providecommand \citenamefont [1]{#1}%
\providecommand \href@noop [0]{\@secondoftwo}%
\providecommand \href [0]{\begingroup \@sanitize@url \@href}%
\providecommand \@href[1]{\@@startlink{#1}\@@href}%
\providecommand \@@href[1]{\endgroup#1\@@endlink}%
\providecommand \@sanitize@url [0]{\catcode `\\12\catcode `\$12\catcode `\&12\catcode `\#12\catcode `\^12\catcode `\_12\catcode `\%12\relax}%
\providecommand \@@startlink[1]{}%
\providecommand \@@endlink[0]{}%
\providecommand \url  [0]{\begingroup\@sanitize@url \@url }%
\providecommand \@url [1]{\endgroup\@href {#1}{\urlprefix }}%
\providecommand \urlprefix  [0]{URL }%
\providecommand \Eprint [0]{\href }%
\providecommand \doibase [0]{https://doi.org/}%
\providecommand \selectlanguage [0]{\@gobble}%
\providecommand \bibinfo  [0]{\@secondoftwo}%
\providecommand \bibfield  [0]{\@secondoftwo}%
\providecommand \translation [1]{[#1]}%
\providecommand \BibitemOpen [0]{}%
\providecommand \bibitemStop [0]{}%
\providecommand \bibitemNoStop [0]{.\EOS\space}%
\providecommand \EOS [0]{\spacefactor3000\relax}%
\providecommand \BibitemShut  [1]{\csname bibitem#1\endcsname}%
\let\auto@bib@innerbib\@empty
\bibitem [{\citenamefont {Sourjik}\ and\ \citenamefont {Wingreen}(2012)}]{sourjik2012responding}%
  \BibitemOpen
  \bibfield  {author} {\bibinfo {author} {\bibfnamefont {V.}~\bibnamefont {Sourjik}}\ and\ \bibinfo {author} {\bibfnamefont {N.~S.}\ \bibnamefont {Wingreen}},\ }\href@noop {} {\bibfield  {journal} {\bibinfo  {journal} {Current opinion in cell biology}\ }\textbf {\bibinfo {volume} {24}},\ \bibinfo {pages} {262} (\bibinfo {year} {2012})}\BibitemShut {NoStop}%
\bibitem [{\citenamefont {Navlakha}\ and\ \citenamefont {Bar-Joseph}(2011)}]{navlakha2011algorithms}%
  \BibitemOpen
  \bibfield  {author} {\bibinfo {author} {\bibfnamefont {S.}~\bibnamefont {Navlakha}}\ and\ \bibinfo {author} {\bibfnamefont {Z.}~\bibnamefont {Bar-Joseph}},\ }\href@noop {} {\bibfield  {journal} {\bibinfo  {journal} {Molecular systems biology}\ }\textbf {\bibinfo {volume} {7}},\ \bibinfo {pages} {546} (\bibinfo {year} {2011})}\BibitemShut {NoStop}%
\bibitem [{\citenamefont {Wolpert}\ and\ \citenamefont {Harper}(2024)}]{wolpert2024computational}%
  \BibitemOpen
  \bibfield  {author} {\bibinfo {author} {\bibfnamefont {D.~H.}\ \bibnamefont {Wolpert}}\ and\ \bibinfo {author} {\bibfnamefont {K.}~\bibnamefont {Harper}},\ }\href@noop {} {\bibfield  {journal} {\bibinfo  {journal} {arXiv preprint arXiv:2408.08861}\ } (\bibinfo {year} {2024})}\BibitemShut {NoStop}%
\bibitem [{\citenamefont {Sipser}(1996)}]{sipser1996introduction}%
  \BibitemOpen
  \bibfield  {author} {\bibinfo {author} {\bibfnamefont {M.}~\bibnamefont {Sipser}},\ }\href@noop {} {\bibfield  {journal} {\bibinfo  {journal} {ACM Sigact News}\ }\textbf {\bibinfo {volume} {27}},\ \bibinfo {pages} {27} (\bibinfo {year} {1996})}\BibitemShut {NoStop}%
\bibitem [{\citenamefont {Arora}\ and\ \citenamefont {Barak}(2009)}]{arora2009computational}%
  \BibitemOpen
  \bibfield  {author} {\bibinfo {author} {\bibfnamefont {S.}~\bibnamefont {Arora}}\ and\ \bibinfo {author} {\bibfnamefont {B.}~\bibnamefont {Barak}},\ }\href@noop {} {\emph {\bibinfo {title} {Computational complexity: a modern approach}}}\ (\bibinfo  {publisher} {Cambridge University Press},\ \bibinfo {year} {2009})\BibitemShut {NoStop}%
\bibitem [{\citenamefont {Li}\ and\ \citenamefont {Vitanyi}(2008)}]{li2008introduction}%
  \BibitemOpen
  \bibfield  {author} {\bibinfo {author} {\bibfnamefont {M.}~\bibnamefont {Li}}\ and\ \bibinfo {author} {\bibfnamefont {P.}~\bibnamefont {Vitanyi}},\ }\href@noop {} {\bibinfo {title} {An introduction to kolmogorov complexity and its applications}} (\bibinfo {year} {2008})\BibitemShut {NoStop}%
\bibitem [{\citenamefont {Piccinini}\ and\ \citenamefont {Maley}(2010)}]{piccinini2010computation}%
  \BibitemOpen
  \bibfield  {author} {\bibinfo {author} {\bibfnamefont {G.}~\bibnamefont {Piccinini}}\ and\ \bibinfo {author} {\bibfnamefont {C.}~\bibnamefont {Maley}},\ }\href@noop {} {\bibinfo {title} {Computation in physical systems}} (\bibinfo {year} {2010})\BibitemShut {NoStop}%
\bibitem [{\citenamefont {Urai}\ \emph {et~al.}(2022)\citenamefont {Urai}, \citenamefont {Doiron}, \citenamefont {Leifer},\ and\ \citenamefont {Churchland}}]{urai2022large}%
  \BibitemOpen
  \bibfield  {author} {\bibinfo {author} {\bibfnamefont {A.~E.}\ \bibnamefont {Urai}}, \bibinfo {author} {\bibfnamefont {B.}~\bibnamefont {Doiron}}, \bibinfo {author} {\bibfnamefont {A.~M.}\ \bibnamefont {Leifer}},\ and\ \bibinfo {author} {\bibfnamefont {A.~K.}\ \bibnamefont {Churchland}},\ }\href@noop {} {\bibfield  {journal} {\bibinfo  {journal} {Nature neuroscience}\ }\textbf {\bibinfo {volume} {25}},\ \bibinfo {pages} {11} (\bibinfo {year} {2022})}\BibitemShut {NoStop}%
\bibitem [{\citenamefont {Sun}\ \emph {et~al.}(2022)\citenamefont {Sun}, \citenamefont {O’Shea}, \citenamefont {Golub}, \citenamefont {Trautmann}, \citenamefont {Vyas}, \citenamefont {Ryu},\ and\ \citenamefont {Shenoy}}]{sun2022cortical}%
  \BibitemOpen
  \bibfield  {author} {\bibinfo {author} {\bibfnamefont {X.}~\bibnamefont {Sun}}, \bibinfo {author} {\bibfnamefont {D.~J.}\ \bibnamefont {O’Shea}}, \bibinfo {author} {\bibfnamefont {M.~D.}\ \bibnamefont {Golub}}, \bibinfo {author} {\bibfnamefont {E.~M.}\ \bibnamefont {Trautmann}}, \bibinfo {author} {\bibfnamefont {S.}~\bibnamefont {Vyas}}, \bibinfo {author} {\bibfnamefont {S.~I.}\ \bibnamefont {Ryu}},\ and\ \bibinfo {author} {\bibfnamefont {K.~V.}\ \bibnamefont {Shenoy}},\ }\href@noop {} {\bibfield  {journal} {\bibinfo  {journal} {Nature}\ }\textbf {\bibinfo {volume} {602}},\ \bibinfo {pages} {274} (\bibinfo {year} {2022})}\BibitemShut {NoStop}%
\bibitem [{\citenamefont {El~Hady}\ \emph {et~al.}(2024)\citenamefont {El~Hady}, \citenamefont {Takahashi}, \citenamefont {Sun}, \citenamefont {Akinwale}, \citenamefont {Boyd-Meredith}, \citenamefont {Zhang}, \citenamefont {Charles},\ and\ \citenamefont {Brody}}]{el2024chronic}%
  \BibitemOpen
  \bibfield  {author} {\bibinfo {author} {\bibfnamefont {A.}~\bibnamefont {El~Hady}}, \bibinfo {author} {\bibfnamefont {D.}~\bibnamefont {Takahashi}}, \bibinfo {author} {\bibfnamefont {R.}~\bibnamefont {Sun}}, \bibinfo {author} {\bibfnamefont {O.}~\bibnamefont {Akinwale}}, \bibinfo {author} {\bibfnamefont {T.}~\bibnamefont {Boyd-Meredith}}, \bibinfo {author} {\bibfnamefont {Y.}~\bibnamefont {Zhang}}, \bibinfo {author} {\bibfnamefont {A.~S.}\ \bibnamefont {Charles}},\ and\ \bibinfo {author} {\bibfnamefont {C.~D.}\ \bibnamefont {Brody}},\ }\href@noop {} {\bibfield  {journal} {\bibinfo  {journal} {Journal of neuroscience methods}\ }\textbf {\bibinfo {volume} {403}},\ \bibinfo {pages} {110033} (\bibinfo {year} {2024})}\BibitemShut {NoStop}%
\bibitem [{\citenamefont {Kim}\ \emph {et~al.}(2017)\citenamefont {Kim}, \citenamefont {Rouault}, \citenamefont {Druckmann},\ and\ \citenamefont {Jayaraman}}]{kim2017ring}%
  \BibitemOpen
  \bibfield  {author} {\bibinfo {author} {\bibfnamefont {S.~S.}\ \bibnamefont {Kim}}, \bibinfo {author} {\bibfnamefont {H.}~\bibnamefont {Rouault}}, \bibinfo {author} {\bibfnamefont {S.}~\bibnamefont {Druckmann}},\ and\ \bibinfo {author} {\bibfnamefont {V.}~\bibnamefont {Jayaraman}},\ }\href@noop {} {\bibfield  {journal} {\bibinfo  {journal} {Science}\ }\textbf {\bibinfo {volume} {356}},\ \bibinfo {pages} {849} (\bibinfo {year} {2017})}\BibitemShut {NoStop}%
\bibitem [{\citenamefont {Jolliffe}\ and\ \citenamefont {Cadima}(2016)}]{jolliffe2016principal}%
  \BibitemOpen
  \bibfield  {author} {\bibinfo {author} {\bibfnamefont {I.~T.}\ \bibnamefont {Jolliffe}}\ and\ \bibinfo {author} {\bibfnamefont {J.}~\bibnamefont {Cadima}},\ }\href@noop {} {\bibfield  {journal} {\bibinfo  {journal} {Philosophical transactions of the royal society A: Mathematical, Physical and Engineering Sciences}\ }\textbf {\bibinfo {volume} {374}},\ \bibinfo {pages} {20150202} (\bibinfo {year} {2016})}\BibitemShut {NoStop}%
\bibitem [{\citenamefont {Kingma}(2013)}]{kingma2013auto}%
  \BibitemOpen
  \bibfield  {author} {\bibinfo {author} {\bibfnamefont {D.~P.}\ \bibnamefont {Kingma}},\ }\href@noop {} {\bibfield  {journal} {\bibinfo  {journal} {arXiv preprint arXiv:1312.6114}\ } (\bibinfo {year} {2013})}\BibitemShut {NoStop}%
\bibitem [{\citenamefont {Box}\ \emph {et~al.}(2015)\citenamefont {Box}, \citenamefont {Jenkins}, \citenamefont {Reinsel},\ and\ \citenamefont {Ljung}}]{box2015time}%
  \BibitemOpen
  \bibfield  {author} {\bibinfo {author} {\bibfnamefont {G.~E.}\ \bibnamefont {Box}}, \bibinfo {author} {\bibfnamefont {G.~M.}\ \bibnamefont {Jenkins}}, \bibinfo {author} {\bibfnamefont {G.~C.}\ \bibnamefont {Reinsel}},\ and\ \bibinfo {author} {\bibfnamefont {G.~M.}\ \bibnamefont {Ljung}},\ }\href@noop {} {\emph {\bibinfo {title} {Time series analysis: forecasting and control}}}\ (\bibinfo  {publisher} {John Wiley \& Sons},\ \bibinfo {year} {2015})\BibitemShut {NoStop}%
\bibitem [{\citenamefont {Li}\ \emph {et~al.}(2020)\citenamefont {Li}, \citenamefont {Wong}, \citenamefont {Chen},\ and\ \citenamefont {Duvenaud}}]{li2020scalable}%
  \BibitemOpen
  \bibfield  {author} {\bibinfo {author} {\bibfnamefont {X.}~\bibnamefont {Li}}, \bibinfo {author} {\bibfnamefont {T.-K.~L.}\ \bibnamefont {Wong}}, \bibinfo {author} {\bibfnamefont {R.~T.~Q.}\ \bibnamefont {Chen}},\ and\ \bibinfo {author} {\bibfnamefont {D.}~\bibnamefont {Duvenaud}},\ }\href@noop {} {\bibfield  {journal} {\bibinfo  {journal} {International Conference on Artificial Intelligence and Statistics}\ } (\bibinfo {year} {2020})}\BibitemShut {NoStop}%
\bibitem [{\citenamefont {Kidger}\ \emph {et~al.}(2021)\citenamefont {Kidger}, \citenamefont {Foster}, \citenamefont {Li}, \citenamefont {Oberhauser},\ and\ \citenamefont {Lyons}}]{kidger2021neuralsde}%
  \BibitemOpen
  \bibfield  {author} {\bibinfo {author} {\bibfnamefont {P.}~\bibnamefont {Kidger}}, \bibinfo {author} {\bibfnamefont {J.}~\bibnamefont {Foster}}, \bibinfo {author} {\bibfnamefont {X.}~\bibnamefont {Li}}, \bibinfo {author} {\bibfnamefont {H.}~\bibnamefont {Oberhauser}},\ and\ \bibinfo {author} {\bibfnamefont {T.}~\bibnamefont {Lyons}},\ }\href@noop {} {\bibfield  {journal} {\bibinfo  {journal} {International Conference on Machine Learning}\ } (\bibinfo {year} {2021})}\BibitemShut {NoStop}%
\bibitem [{\citenamefont {Doedel}\ \emph {et~al.}(2015)\citenamefont {Doedel}, \citenamefont {Krauskopf},\ and\ \citenamefont {Osinga}}]{doedel2015global}%
  \BibitemOpen
  \bibfield  {author} {\bibinfo {author} {\bibfnamefont {E.~J.}\ \bibnamefont {Doedel}}, \bibinfo {author} {\bibfnamefont {B.}~\bibnamefont {Krauskopf}},\ and\ \bibinfo {author} {\bibfnamefont {H.~M.}\ \bibnamefont {Osinga}},\ }\href@noop {} {\bibfield  {journal} {\bibinfo  {journal} {Nonlinearity}\ }\textbf {\bibinfo {volume} {28}},\ \bibinfo {pages} {R113} (\bibinfo {year} {2015})}\BibitemShut {NoStop}%
\bibitem [{\citenamefont {Wolfram}(1984)}]{wolfram1984cellular}%
  \BibitemOpen
  \bibfield  {author} {\bibinfo {author} {\bibfnamefont {S.}~\bibnamefont {Wolfram}},\ }\href@noop {} {\bibfield  {journal} {\bibinfo  {journal} {Nature}\ }\textbf {\bibinfo {volume} {311}},\ \bibinfo {pages} {419} (\bibinfo {year} {1984})}\BibitemShut {NoStop}%
\bibitem [{\citenamefont {Kaplan}\ \emph {et~al.}(2020)\citenamefont {Kaplan}, \citenamefont {Thula}, \citenamefont {Khoss},\ and\ \citenamefont {Zimmer}}]{kaplan2020nested}%
  \BibitemOpen
  \bibfield  {author} {\bibinfo {author} {\bibfnamefont {H.~S.}\ \bibnamefont {Kaplan}}, \bibinfo {author} {\bibfnamefont {O.~S.}\ \bibnamefont {Thula}}, \bibinfo {author} {\bibfnamefont {N.}~\bibnamefont {Khoss}},\ and\ \bibinfo {author} {\bibfnamefont {M.}~\bibnamefont {Zimmer}},\ }\href@noop {} {\bibfield  {journal} {\bibinfo  {journal} {Neuron}\ }\textbf {\bibinfo {volume} {105}},\ \bibinfo {pages} {562} (\bibinfo {year} {2020})}\BibitemShut {NoStop}%
\bibitem [{\citenamefont {Ji}\ \emph {et~al.}(2021)\citenamefont {Ji}, \citenamefont {Madan}, \citenamefont {Fabre}, \citenamefont {Dayan}, \citenamefont {Baker}, \citenamefont {Kramer}, \citenamefont {Nwabudike},\ and\ \citenamefont {Flavell}}]{ji2021neural}%
  \BibitemOpen
  \bibfield  {author} {\bibinfo {author} {\bibfnamefont {N.}~\bibnamefont {Ji}}, \bibinfo {author} {\bibfnamefont {G.~K.}\ \bibnamefont {Madan}}, \bibinfo {author} {\bibfnamefont {G.~I.}\ \bibnamefont {Fabre}}, \bibinfo {author} {\bibfnamefont {A.}~\bibnamefont {Dayan}}, \bibinfo {author} {\bibfnamefont {C.~M.}\ \bibnamefont {Baker}}, \bibinfo {author} {\bibfnamefont {T.~S.}\ \bibnamefont {Kramer}}, \bibinfo {author} {\bibfnamefont {I.}~\bibnamefont {Nwabudike}},\ and\ \bibinfo {author} {\bibfnamefont {S.~W.}\ \bibnamefont {Flavell}},\ }\href@noop {} {\bibfield  {journal} {\bibinfo  {journal} {Elife}\ }\textbf {\bibinfo {volume} {10}},\ \bibinfo {pages} {e62889} (\bibinfo {year} {2021})}\BibitemShut {NoStop}%
\bibitem [{\citenamefont {Hallinen}\ \emph {et~al.}(2021)\citenamefont {Hallinen}, \citenamefont {Dempsey}, \citenamefont {Scholz}, \citenamefont {Yu}, \citenamefont {Linder}, \citenamefont {Randi}, \citenamefont {Sharma}, \citenamefont {Shaevitz},\ and\ \citenamefont {Leifer}}]{hallinen2021decoding}%
  \BibitemOpen
  \bibfield  {author} {\bibinfo {author} {\bibfnamefont {K.~M.}\ \bibnamefont {Hallinen}}, \bibinfo {author} {\bibfnamefont {R.}~\bibnamefont {Dempsey}}, \bibinfo {author} {\bibfnamefont {M.}~\bibnamefont {Scholz}}, \bibinfo {author} {\bibfnamefont {X.}~\bibnamefont {Yu}}, \bibinfo {author} {\bibfnamefont {A.}~\bibnamefont {Linder}}, \bibinfo {author} {\bibfnamefont {F.}~\bibnamefont {Randi}}, \bibinfo {author} {\bibfnamefont {A.~K.}\ \bibnamefont {Sharma}}, \bibinfo {author} {\bibfnamefont {J.~W.}\ \bibnamefont {Shaevitz}},\ and\ \bibinfo {author} {\bibfnamefont {A.~M.}\ \bibnamefont {Leifer}},\ }\href@noop {} {\bibfield  {journal} {\bibinfo  {journal} {Elife}\ }\textbf {\bibinfo {volume} {10}},\ \bibinfo {pages} {e66135} (\bibinfo {year} {2021})}\BibitemShut {NoStop}%
\bibitem [{\citenamefont {Nguyen}\ \emph {et~al.}(2016)\citenamefont {Nguyen}, \citenamefont {Shipley}, \citenamefont {Linder}, \citenamefont {Plummer}, \citenamefont {Liu}, \citenamefont {Setru}, \citenamefont {Shaevitz},\ and\ \citenamefont {Leifer}}]{nguyen2016whole}%
  \BibitemOpen
  \bibfield  {author} {\bibinfo {author} {\bibfnamefont {J.~P.}\ \bibnamefont {Nguyen}}, \bibinfo {author} {\bibfnamefont {F.~B.}\ \bibnamefont {Shipley}}, \bibinfo {author} {\bibfnamefont {A.~N.}\ \bibnamefont {Linder}}, \bibinfo {author} {\bibfnamefont {G.~S.}\ \bibnamefont {Plummer}}, \bibinfo {author} {\bibfnamefont {M.}~\bibnamefont {Liu}}, \bibinfo {author} {\bibfnamefont {S.~U.}\ \bibnamefont {Setru}}, \bibinfo {author} {\bibfnamefont {J.~W.}\ \bibnamefont {Shaevitz}},\ and\ \bibinfo {author} {\bibfnamefont {A.~M.}\ \bibnamefont {Leifer}},\ }\href@noop {} {\bibfield  {journal} {\bibinfo  {journal} {Proceedings of the National Academy of Sciences}\ }\textbf {\bibinfo {volume} {113}},\ \bibinfo {pages} {E1074} (\bibinfo {year} {2016})}\BibitemShut {NoStop}%
\bibitem [{\citenamefont {Goodfellow}\ \emph {et~al.}(2016)\citenamefont {Goodfellow}, \citenamefont {Bengio}, \citenamefont {Courville},\ and\ \citenamefont {Bengio}}]{goodfellow2016deep}%
  \BibitemOpen
  \bibfield  {author} {\bibinfo {author} {\bibfnamefont {I.}~\bibnamefont {Goodfellow}}, \bibinfo {author} {\bibfnamefont {Y.}~\bibnamefont {Bengio}}, \bibinfo {author} {\bibfnamefont {A.}~\bibnamefont {Courville}},\ and\ \bibinfo {author} {\bibfnamefont {Y.}~\bibnamefont {Bengio}},\ }\href@noop {} {\emph {\bibinfo {title} {Deep learning}}},\ Vol.~\bibinfo {volume} {1}\ (\bibinfo  {publisher} {MIT press Cambridge},\ \bibinfo {year} {2016})\BibitemShut {NoStop}%
\bibitem [{\citenamefont {Cook}\ \emph {et~al.}(2004)\citenamefont {Cook} \emph {et~al.}}]{cook2004universality}%
  \BibitemOpen
  \bibfield  {author} {\bibinfo {author} {\bibfnamefont {M.}~\bibnamefont {Cook}} \emph {et~al.},\ }\href@noop {} {\bibfield  {journal} {\bibinfo  {journal} {Complex systems}\ }\textbf {\bibinfo {volume} {15}},\ \bibinfo {pages} {1} (\bibinfo {year} {2004})}\BibitemShut {NoStop}%
\bibitem [{\citenamefont {Kato}\ \emph {et~al.}(2015)\citenamefont {Kato}, \citenamefont {Kaplan}, \citenamefont {Schr{\"o}del}, \citenamefont {Skora}, \citenamefont {Lindsay}, \citenamefont {Yemini}, \citenamefont {Lockery},\ and\ \citenamefont {Zimmer}}]{kato2015global}%
  \BibitemOpen
  \bibfield  {author} {\bibinfo {author} {\bibfnamefont {S.}~\bibnamefont {Kato}}, \bibinfo {author} {\bibfnamefont {H.~S.}\ \bibnamefont {Kaplan}}, \bibinfo {author} {\bibfnamefont {T.}~\bibnamefont {Schr{\"o}del}}, \bibinfo {author} {\bibfnamefont {S.}~\bibnamefont {Skora}}, \bibinfo {author} {\bibfnamefont {T.~H.}\ \bibnamefont {Lindsay}}, \bibinfo {author} {\bibfnamefont {E.}~\bibnamefont {Yemini}}, \bibinfo {author} {\bibfnamefont {S.}~\bibnamefont {Lockery}},\ and\ \bibinfo {author} {\bibfnamefont {M.}~\bibnamefont {Zimmer}},\ }\href@noop {} {\bibfield  {journal} {\bibinfo  {journal} {Cell}\ }\textbf {\bibinfo {volume} {163}},\ \bibinfo {pages} {656} (\bibinfo {year} {2015})}\BibitemShut {NoStop}%
\bibitem [{\citenamefont {Gauthey}\ \emph {et~al.}(2024)\citenamefont {Gauthey}, \citenamefont {Randi}, \citenamefont {Sharma}, \citenamefont {Kumar},\ and\ \citenamefont {Leifer}}]{gauthey2024light}%
  \BibitemOpen
  \bibfield  {author} {\bibinfo {author} {\bibfnamefont {W.}~\bibnamefont {Gauthey}}, \bibinfo {author} {\bibfnamefont {F.}~\bibnamefont {Randi}}, \bibinfo {author} {\bibfnamefont {A.~K.}\ \bibnamefont {Sharma}}, \bibinfo {author} {\bibfnamefont {S.}~\bibnamefont {Kumar}},\ and\ \bibinfo {author} {\bibfnamefont {A.~M.}\ \bibnamefont {Leifer}},\ }\href@noop {} {\bibfield  {journal} {\bibinfo  {journal} {Current Biology}\ }\textbf {\bibinfo {volume} {34}},\ \bibinfo {pages} {R14} (\bibinfo {year} {2024})}\BibitemShut {NoStop}%
\bibitem [{\citenamefont {Sabrin}\ \emph {et~al.}(2020)\citenamefont {Sabrin}, \citenamefont {Wei}, \citenamefont {van~den Heuvel},\ and\ \citenamefont {Dovrolis}}]{sabrin2020hourglass}%
  \BibitemOpen
  \bibfield  {author} {\bibinfo {author} {\bibfnamefont {K.~M.}\ \bibnamefont {Sabrin}}, \bibinfo {author} {\bibfnamefont {Y.}~\bibnamefont {Wei}}, \bibinfo {author} {\bibfnamefont {M.~P.}\ \bibnamefont {van~den Heuvel}},\ and\ \bibinfo {author} {\bibfnamefont {C.}~\bibnamefont {Dovrolis}},\ }\href@noop {} {\bibfield  {journal} {\bibinfo  {journal} {PLoS computational biology}\ }\textbf {\bibinfo {volume} {16}},\ \bibinfo {pages} {e1007526} (\bibinfo {year} {2020})}\BibitemShut {NoStop}%
\bibitem [{\citenamefont {Yemini}\ \emph {et~al.}(2021)\citenamefont {Yemini}, \citenamefont {Lin}, \citenamefont {Nejatbakhsh}, \citenamefont {Varol}, \citenamefont {Sun}, \citenamefont {Mena}, \citenamefont {Samuel}, \citenamefont {Paninski}, \citenamefont {Venkatachalam},\ and\ \citenamefont {Hobert}}]{yemini2021neuropal}%
  \BibitemOpen
  \bibfield  {author} {\bibinfo {author} {\bibfnamefont {E.}~\bibnamefont {Yemini}}, \bibinfo {author} {\bibfnamefont {A.}~\bibnamefont {Lin}}, \bibinfo {author} {\bibfnamefont {A.}~\bibnamefont {Nejatbakhsh}}, \bibinfo {author} {\bibfnamefont {E.}~\bibnamefont {Varol}}, \bibinfo {author} {\bibfnamefont {R.}~\bibnamefont {Sun}}, \bibinfo {author} {\bibfnamefont {G.~E.}\ \bibnamefont {Mena}}, \bibinfo {author} {\bibfnamefont {A.~D.}\ \bibnamefont {Samuel}}, \bibinfo {author} {\bibfnamefont {L.}~\bibnamefont {Paninski}}, \bibinfo {author} {\bibfnamefont {V.}~\bibnamefont {Venkatachalam}},\ and\ \bibinfo {author} {\bibfnamefont {O.}~\bibnamefont {Hobert}},\ }\href@noop {} {\bibfield  {journal} {\bibinfo  {journal} {Cell}\ }\textbf {\bibinfo {volume} {184}},\ \bibinfo {pages} {272} (\bibinfo {year} {2021})}\BibitemShut {NoStop}%
\bibitem [{\citenamefont {Atanas}\ \emph {et~al.}(2023)\citenamefont {Atanas}, \citenamefont {Kim1}, \citenamefont {Wang}, \citenamefont {Bueno}, \citenamefont {Becker}, \citenamefont {Kang}, \citenamefont {Park}, \citenamefont {Kramer}, \citenamefont {Wan}, \citenamefont {Baskoylu}, \citenamefont {Dag}, \citenamefont {Kalogeropoulou}, \citenamefont {Gomes}, \citenamefont {Estrem}, \citenamefont {Cohen}, \citenamefont {Mansinghka},\ and\ \citenamefont {Flavell}}]{atanas2023brain}%
  \BibitemOpen
  \bibfield  {author} {\bibinfo {author} {\bibfnamefont {A.~A.}\ \bibnamefont {Atanas}}, \bibinfo {author} {\bibfnamefont {J.}~\bibnamefont {Kim1}}, \bibinfo {author} {\bibfnamefont {Z.}~\bibnamefont {Wang}}, \bibinfo {author} {\bibfnamefont {E.}~\bibnamefont {Bueno}}, \bibinfo {author} {\bibfnamefont {M.}~\bibnamefont {Becker}}, \bibinfo {author} {\bibfnamefont {D.}~\bibnamefont {Kang}}, \bibinfo {author} {\bibfnamefont {J.}~\bibnamefont {Park}}, \bibinfo {author} {\bibfnamefont {T.~S.}\ \bibnamefont {Kramer}}, \bibinfo {author} {\bibfnamefont {F.~K.}\ \bibnamefont {Wan}}, \bibinfo {author} {\bibfnamefont {S.}~\bibnamefont {Baskoylu}}, \bibinfo {author} {\bibfnamefont {U.}~\bibnamefont {Dag}}, \bibinfo {author} {\bibfnamefont {E.}~\bibnamefont {Kalogeropoulou}}, \bibinfo {author} {\bibfnamefont {M.~A.}\ \bibnamefont {Gomes}}, \bibinfo {author} {\bibfnamefont {C.}~\bibnamefont {Estrem}}, \bibinfo {author} {\bibfnamefont {N.}~\bibnamefont {Cohen}}, \bibinfo {author} {\bibfnamefont {V.~K.}\ \bibnamefont
  {Mansinghka}},\ and\ \bibinfo {author} {\bibfnamefont {S.~W.}\ \bibnamefont {Flavell}},\ }\href {https://doi.org/10.1016/j.cell.2023.08.033} {\bibfield  {journal} {\bibinfo  {journal} {Cell}\ }\textbf {\bibinfo {volume} {186}},\ \bibinfo {pages} {4134} (\bibinfo {year} {2023})}\BibitemShut {NoStop}%
\bibitem [{\citenamefont {O’Shea}\ \emph {et~al.}(2022)\citenamefont {O’Shea}, \citenamefont {Duncker}, \citenamefont {Goo}, \citenamefont {Sun}, \citenamefont {Vyas}, \citenamefont {Trautmann}, \citenamefont {Diester}, \citenamefont {Ramakrishnan}, \citenamefont {Deisseroth}, \citenamefont {Sahani} \emph {et~al.}}]{o2022direct}%
  \BibitemOpen
  \bibfield  {author} {\bibinfo {author} {\bibfnamefont {D.~J.}\ \bibnamefont {O’Shea}}, \bibinfo {author} {\bibfnamefont {L.}~\bibnamefont {Duncker}}, \bibinfo {author} {\bibfnamefont {W.}~\bibnamefont {Goo}}, \bibinfo {author} {\bibfnamefont {X.}~\bibnamefont {Sun}}, \bibinfo {author} {\bibfnamefont {S.}~\bibnamefont {Vyas}}, \bibinfo {author} {\bibfnamefont {E.~M.}\ \bibnamefont {Trautmann}}, \bibinfo {author} {\bibfnamefont {I.}~\bibnamefont {Diester}}, \bibinfo {author} {\bibfnamefont {C.}~\bibnamefont {Ramakrishnan}}, \bibinfo {author} {\bibfnamefont {K.}~\bibnamefont {Deisseroth}}, \bibinfo {author} {\bibfnamefont {M.}~\bibnamefont {Sahani}}, \emph {et~al.},\ }\href@noop {} {\bibfield  {journal} {\bibinfo  {journal} {bioRxiv}\ ,\ \bibinfo {pages} {2022}} (\bibinfo {year} {2022})}\BibitemShut {NoStop}%
\bibitem [{\citenamefont {Krakauer}\ \emph {et~al.}(2017)\citenamefont {Krakauer}, \citenamefont {Ghazanfar}, \citenamefont {Gomez-Marin}, \citenamefont {MacIver},\ and\ \citenamefont {Poeppel}}]{krakauer2017neuroscience}%
  \BibitemOpen
  \bibfield  {author} {\bibinfo {author} {\bibfnamefont {J.~W.}\ \bibnamefont {Krakauer}}, \bibinfo {author} {\bibfnamefont {A.~A.}\ \bibnamefont {Ghazanfar}}, \bibinfo {author} {\bibfnamefont {A.}~\bibnamefont {Gomez-Marin}}, \bibinfo {author} {\bibfnamefont {M.~A.}\ \bibnamefont {MacIver}},\ and\ \bibinfo {author} {\bibfnamefont {D.}~\bibnamefont {Poeppel}},\ }\href@noop {} {\bibfield  {journal} {\bibinfo  {journal} {Neuron}\ }\textbf {\bibinfo {volume} {93}},\ \bibinfo {pages} {480} (\bibinfo {year} {2017})}\BibitemShut {NoStop}%
\bibitem [{\citenamefont {Datta}\ \emph {et~al.}(2019)\citenamefont {Datta}, \citenamefont {Anderson}, \citenamefont {Branson}, \citenamefont {Perona},\ and\ \citenamefont {Leifer}}]{datta2019computational}%
  \BibitemOpen
  \bibfield  {author} {\bibinfo {author} {\bibfnamefont {S.~R.}\ \bibnamefont {Datta}}, \bibinfo {author} {\bibfnamefont {D.~J.}\ \bibnamefont {Anderson}}, \bibinfo {author} {\bibfnamefont {K.}~\bibnamefont {Branson}}, \bibinfo {author} {\bibfnamefont {P.}~\bibnamefont {Perona}},\ and\ \bibinfo {author} {\bibfnamefont {A.}~\bibnamefont {Leifer}},\ }\href@noop {} {\bibfield  {journal} {\bibinfo  {journal} {Neuron}\ }\textbf {\bibinfo {volume} {104}},\ \bibinfo {pages} {11} (\bibinfo {year} {2019})}\BibitemShut {NoStop}%
\bibitem [{\citenamefont {Van~Essen}\ \emph {et~al.}(2013)\citenamefont {Van~Essen}, \citenamefont {Smith}, \citenamefont {Barch}, \citenamefont {Behrens}, \citenamefont {Yacoub}, \citenamefont {Ugurbil}, \citenamefont {Consortium} \emph {et~al.}}]{van2013wu}%
  \BibitemOpen
  \bibfield  {author} {\bibinfo {author} {\bibfnamefont {D.~C.}\ \bibnamefont {Van~Essen}}, \bibinfo {author} {\bibfnamefont {S.~M.}\ \bibnamefont {Smith}}, \bibinfo {author} {\bibfnamefont {D.~M.}\ \bibnamefont {Barch}}, \bibinfo {author} {\bibfnamefont {T.~E.}\ \bibnamefont {Behrens}}, \bibinfo {author} {\bibfnamefont {E.}~\bibnamefont {Yacoub}}, \bibinfo {author} {\bibfnamefont {K.}~\bibnamefont {Ugurbil}}, \bibinfo {author} {\bibfnamefont {W.-M.~H.}\ \bibnamefont {Consortium}}, \emph {et~al.},\ }\href@noop {} {\bibfield  {journal} {\bibinfo  {journal} {Neuroimage}\ }\textbf {\bibinfo {volume} {80}},\ \bibinfo {pages} {62} (\bibinfo {year} {2013})}\BibitemShut {NoStop}%
\bibitem [{\citenamefont {Dorkenwald}\ \emph {et~al.}(2024)\citenamefont {Dorkenwald}, \citenamefont {Matsliah}, \citenamefont {Sterling}, \citenamefont {Schlegel}, \citenamefont {chieh Yu}, \citenamefont {McKellar}, \citenamefont {Lin}, \citenamefont {Costa}, \citenamefont {Eichler}, \citenamefont {Yin}, \citenamefont {Silversmith}, \citenamefont {Schneider-Mizell}, \citenamefont {Jordan}, \citenamefont {Brittain}, \citenamefont {Halageri}, \citenamefont {Kuehner}, \citenamefont {Ogedengbe}, \citenamefont {Morey}, \citenamefont {Gager}, \citenamefont {Kruk}, \citenamefont {Perlman}, \citenamefont {Yang}, \citenamefont {Deutsch}, \citenamefont {Bland}, \citenamefont {Sorek}, \citenamefont {Lu}, \citenamefont {Macrina}, \citenamefont {Lee}, \citenamefont {Bae}, \citenamefont {Mu}, \citenamefont {Nehoran}, \citenamefont {Mitchell}, \citenamefont {Popovych}, \citenamefont {Wu}, \citenamefont {Jia}, \citenamefont {Castro}, \citenamefont {Kemnitz}, \citenamefont {Ih}, \citenamefont {Bates}, \citenamefont
  {Eckstein}, \citenamefont {Funke}, \citenamefont {Collman}, \citenamefont {Bock}, \citenamefont {Jefferis}, \citenamefont {Seung}, \citenamefont {Murthy},\ and\ \citenamefont {Consortium}}]{Dorkenwald2024}%
  \BibitemOpen
  \bibfield  {author} {\bibinfo {author} {\bibfnamefont {S.}~\bibnamefont {Dorkenwald}}, \bibinfo {author} {\bibfnamefont {A.}~\bibnamefont {Matsliah}}, \bibinfo {author} {\bibfnamefont {A.~R.}\ \bibnamefont {Sterling}}, \bibinfo {author} {\bibfnamefont {P.}~\bibnamefont {Schlegel}}, \bibinfo {author} {\bibfnamefont {S.}~\bibnamefont {chieh Yu}}, \bibinfo {author} {\bibfnamefont {C.~E.}\ \bibnamefont {McKellar}}, \bibinfo {author} {\bibfnamefont {A.}~\bibnamefont {Lin}}, \bibinfo {author} {\bibfnamefont {M.}~\bibnamefont {Costa}}, \bibinfo {author} {\bibfnamefont {K.}~\bibnamefont {Eichler}}, \bibinfo {author} {\bibfnamefont {Y.}~\bibnamefont {Yin}}, \bibinfo {author} {\bibfnamefont {W.}~\bibnamefont {Silversmith}}, \bibinfo {author} {\bibfnamefont {C.}~\bibnamefont {Schneider-Mizell}}, \bibinfo {author} {\bibfnamefont {C.~S.}\ \bibnamefont {Jordan}}, \bibinfo {author} {\bibfnamefont {D.}~\bibnamefont {Brittain}}, \bibinfo {author} {\bibfnamefont {A.}~\bibnamefont {Halageri}}, \bibinfo {author} {\bibfnamefont
  {K.}~\bibnamefont {Kuehner}}, \bibinfo {author} {\bibfnamefont {O.}~\bibnamefont {Ogedengbe}}, \bibinfo {author} {\bibfnamefont {R.}~\bibnamefont {Morey}}, \bibinfo {author} {\bibfnamefont {J.}~\bibnamefont {Gager}}, \bibinfo {author} {\bibfnamefont {K.}~\bibnamefont {Kruk}}, \bibinfo {author} {\bibfnamefont {E.}~\bibnamefont {Perlman}}, \bibinfo {author} {\bibfnamefont {R.}~\bibnamefont {Yang}}, \bibinfo {author} {\bibfnamefont {D.}~\bibnamefont {Deutsch}}, \bibinfo {author} {\bibfnamefont {D.}~\bibnamefont {Bland}}, \bibinfo {author} {\bibfnamefont {M.}~\bibnamefont {Sorek}}, \bibinfo {author} {\bibfnamefont {R.}~\bibnamefont {Lu}}, \bibinfo {author} {\bibfnamefont {T.}~\bibnamefont {Macrina}}, \bibinfo {author} {\bibfnamefont {K.}~\bibnamefont {Lee}}, \bibinfo {author} {\bibfnamefont {J.~A.}\ \bibnamefont {Bae}}, \bibinfo {author} {\bibfnamefont {S.}~\bibnamefont {Mu}}, \bibinfo {author} {\bibfnamefont {B.}~\bibnamefont {Nehoran}}, \bibinfo {author} {\bibfnamefont {E.}~\bibnamefont {Mitchell}}, \bibinfo
  {author} {\bibfnamefont {S.}~\bibnamefont {Popovych}}, \bibinfo {author} {\bibfnamefont {J.}~\bibnamefont {Wu}}, \bibinfo {author} {\bibfnamefont {Z.}~\bibnamefont {Jia}}, \bibinfo {author} {\bibfnamefont {M.~A.}\ \bibnamefont {Castro}}, \bibinfo {author} {\bibfnamefont {N.}~\bibnamefont {Kemnitz}}, \bibinfo {author} {\bibfnamefont {D.}~\bibnamefont {Ih}}, \bibinfo {author} {\bibfnamefont {A.~S.}\ \bibnamefont {Bates}}, \bibinfo {author} {\bibfnamefont {N.}~\bibnamefont {Eckstein}}, \bibinfo {author} {\bibfnamefont {J.}~\bibnamefont {Funke}}, \bibinfo {author} {\bibfnamefont {F.}~\bibnamefont {Collman}}, \bibinfo {author} {\bibfnamefont {D.~D.}\ \bibnamefont {Bock}}, \bibinfo {author} {\bibfnamefont {G.~S. X.~E.}\ \bibnamefont {Jefferis}}, \bibinfo {author} {\bibfnamefont {H.~S.}\ \bibnamefont {Seung}}, \bibinfo {author} {\bibfnamefont {M.}~\bibnamefont {Murthy}},\ and\ \bibinfo {author} {\bibfnamefont {T.~F.}\ \bibnamefont {Consortium}},\ }\href {https://doi.org/10.1038/s41586-024-07558-y} {\bibfield
  {journal} {\bibinfo  {journal} {Nature}\ }\textbf {\bibinfo {volume} {634}},\ \bibinfo {pages} {124} (\bibinfo {year} {2024})}\BibitemShut {NoStop}%
\bibitem [{\citenamefont {Wolpert}\ \emph {et~al.}(2023)\citenamefont {Wolpert}, \citenamefont {Korbel}, \citenamefont {Lynn}, \citenamefont {Tasnim}, \citenamefont {Grochow}, \citenamefont {Karde{\c{s}}}, \citenamefont {Aimone}, \citenamefont {Balasubramanian}, \citenamefont {De~Giuli}, \citenamefont {Doty} \emph {et~al.}}]{wolpert2023stochastic}%
  \BibitemOpen
  \bibfield  {author} {\bibinfo {author} {\bibfnamefont {D.}~\bibnamefont {Wolpert}}, \bibinfo {author} {\bibfnamefont {J.}~\bibnamefont {Korbel}}, \bibinfo {author} {\bibfnamefont {C.}~\bibnamefont {Lynn}}, \bibinfo {author} {\bibfnamefont {F.}~\bibnamefont {Tasnim}}, \bibinfo {author} {\bibfnamefont {J.}~\bibnamefont {Grochow}}, \bibinfo {author} {\bibfnamefont {G.}~\bibnamefont {Karde{\c{s}}}}, \bibinfo {author} {\bibfnamefont {J.}~\bibnamefont {Aimone}}, \bibinfo {author} {\bibfnamefont {V.}~\bibnamefont {Balasubramanian}}, \bibinfo {author} {\bibfnamefont {E.}~\bibnamefont {De~Giuli}}, \bibinfo {author} {\bibfnamefont {D.}~\bibnamefont {Doty}}, \emph {et~al.},\ }\href@noop {} {\bibfield  {journal} {\bibinfo  {journal} {arXiv preprint arXiv:2311.17166}\ } (\bibinfo {year} {2023})}\BibitemShut {NoStop}%
\bibitem [{\citenamefont {Sol{\'e}}\ \emph {et~al.}(2024)\citenamefont {Sol{\'e}}, \citenamefont {Kempes}, \citenamefont {Corominas-Murtra}, \citenamefont {De~Domenico}, \citenamefont {Kolchinsky}, \citenamefont {Lachmann}, \citenamefont {Libby}, \citenamefont {Saavedra}, \citenamefont {Smith},\ and\ \citenamefont {Wolpert}}]{sole2024fundamental}%
  \BibitemOpen
  \bibfield  {author} {\bibinfo {author} {\bibfnamefont {R.}~\bibnamefont {Sol{\'e}}}, \bibinfo {author} {\bibfnamefont {C.~P.}\ \bibnamefont {Kempes}}, \bibinfo {author} {\bibfnamefont {B.}~\bibnamefont {Corominas-Murtra}}, \bibinfo {author} {\bibfnamefont {M.}~\bibnamefont {De~Domenico}}, \bibinfo {author} {\bibfnamefont {A.}~\bibnamefont {Kolchinsky}}, \bibinfo {author} {\bibfnamefont {M.}~\bibnamefont {Lachmann}}, \bibinfo {author} {\bibfnamefont {E.}~\bibnamefont {Libby}}, \bibinfo {author} {\bibfnamefont {S.}~\bibnamefont {Saavedra}}, \bibinfo {author} {\bibfnamefont {E.}~\bibnamefont {Smith}},\ and\ \bibinfo {author} {\bibfnamefont {D.}~\bibnamefont {Wolpert}},\ }\href@noop {} {\bibfield  {journal} {\bibinfo  {journal} {Interface Focus}\ }\textbf {\bibinfo {volume} {14}},\ \bibinfo {pages} {20240010} (\bibinfo {year} {2024})}\BibitemShut {NoStop}%
\bibitem [{\citenamefont {Wolpert}\ \emph {et~al.}(2024)\citenamefont {Wolpert}, \citenamefont {Korbel},\ and\ \citenamefont {Hofer}}]{wolpert2024focus}%
  \BibitemOpen
  \bibfield  {author} {\bibinfo {author} {\bibfnamefont {D.}~\bibnamefont {Wolpert}}, \bibinfo {author} {\bibfnamefont {J.}~\bibnamefont {Korbel}},\ and\ \bibinfo {author} {\bibfnamefont {M.}~\bibnamefont {Hofer}},\ }\href@noop {} {\bibfield  {journal} {\bibinfo  {journal} {Journal of Physics: Complexity}\ } (\bibinfo {year} {2024})}\BibitemShut {NoStop}%
\bibitem [{\citenamefont {Pereira}\ \emph {et~al.}(2022)\citenamefont {Pereira}, \citenamefont {Tabris}, \citenamefont {Matsliah}, \citenamefont {Turner}, \citenamefont {Li}, \citenamefont {Ravindranath}, \citenamefont {Papadoyannis}, \citenamefont {Normand}, \citenamefont {Deutsch}, \citenamefont {Wang} \emph {et~al.}}]{pereira2022sleap}%
  \BibitemOpen
  \bibfield  {author} {\bibinfo {author} {\bibfnamefont {T.~D.}\ \bibnamefont {Pereira}}, \bibinfo {author} {\bibfnamefont {N.}~\bibnamefont {Tabris}}, \bibinfo {author} {\bibfnamefont {A.}~\bibnamefont {Matsliah}}, \bibinfo {author} {\bibfnamefont {D.~M.}\ \bibnamefont {Turner}}, \bibinfo {author} {\bibfnamefont {J.}~\bibnamefont {Li}}, \bibinfo {author} {\bibfnamefont {S.}~\bibnamefont {Ravindranath}}, \bibinfo {author} {\bibfnamefont {E.~S.}\ \bibnamefont {Papadoyannis}}, \bibinfo {author} {\bibfnamefont {E.}~\bibnamefont {Normand}}, \bibinfo {author} {\bibfnamefont {D.~S.}\ \bibnamefont {Deutsch}}, \bibinfo {author} {\bibfnamefont {Z.~Y.}\ \bibnamefont {Wang}}, \emph {et~al.},\ }\href@noop {} {\bibfield  {journal} {\bibinfo  {journal} {Nature methods}\ }\textbf {\bibinfo {volume} {19}},\ \bibinfo {pages} {486} (\bibinfo {year} {2022})}\BibitemShut {NoStop}%
\bibitem [{\citenamefont {Nguyen}\ \emph {et~al.}(2017)\citenamefont {Nguyen}, \citenamefont {Linder}, \citenamefont {Plummer}, \citenamefont {Shaevitz},\ and\ \citenamefont {Leifer}}]{Nguyen2017}%
  \BibitemOpen
  \bibfield  {author} {\bibinfo {author} {\bibfnamefont {J.~P.}\ \bibnamefont {Nguyen}}, \bibinfo {author} {\bibfnamefont {A.~N.}\ \bibnamefont {Linder}}, \bibinfo {author} {\bibfnamefont {G.~S.}\ \bibnamefont {Plummer}}, \bibinfo {author} {\bibfnamefont {J.~W.}\ \bibnamefont {Shaevitz}},\ and\ \bibinfo {author} {\bibfnamefont {A.~M.}\ \bibnamefont {Leifer}},\ }\href {https://doi.org/10.1371/journal.pcbi.1005517} {\bibfield  {journal} {\bibinfo  {journal} {PLOS Computational Biology}\ }\textbf {\bibinfo {volume} {13}},\ \bibinfo {pages} {e1005517} (\bibinfo {year} {2017})}\BibitemShut {NoStop}%
\bibitem [{\citenamefont {Creamer}\ \emph {et~al.}(2022)\citenamefont {Creamer}, \citenamefont {Chen}, \citenamefont {Leifer},\ and\ \citenamefont {Pillow}}]{Creamer2022}%
  \BibitemOpen
  \bibfield  {author} {\bibinfo {author} {\bibfnamefont {M.~S.}\ \bibnamefont {Creamer}}, \bibinfo {author} {\bibfnamefont {K.~S.}\ \bibnamefont {Chen}}, \bibinfo {author} {\bibfnamefont {A.~M.}\ \bibnamefont {Leifer}},\ and\ \bibinfo {author} {\bibfnamefont {J.~W.}\ \bibnamefont {Pillow}},\ }\href {https://doi.org/10.1371/journal.pcbi.1010421} {\bibfield  {journal} {\bibinfo  {journal} {PLOS Computational Biology}\ }\textbf {\bibinfo {volume} {18}},\ \bibinfo {pages} {e1010421} (\bibinfo {year} {2022})}\BibitemShut {NoStop}%
\bibitem [{\citenamefont {Stephens}\ \emph {et~al.}(2008)\citenamefont {Stephens}, \citenamefont {Johnson-Kerner}, \citenamefont {Bialek},\ and\ \citenamefont {Ryu}}]{Stephens2008}%
  \BibitemOpen
  \bibfield  {author} {\bibinfo {author} {\bibfnamefont {G.~J.}\ \bibnamefont {Stephens}}, \bibinfo {author} {\bibfnamefont {B.}~\bibnamefont {Johnson-Kerner}}, \bibinfo {author} {\bibfnamefont {W.}~\bibnamefont {Bialek}},\ and\ \bibinfo {author} {\bibfnamefont {W.~S.}\ \bibnamefont {Ryu}},\ }\href {https://doi.org/10.1371/journal.pcbi.1000028} {\bibfield  {journal} {\bibinfo  {journal} {PLOS Computational Biology}\ }\textbf {\bibinfo {volume} {4}},\ \bibinfo {pages} {e1000028} (\bibinfo {year} {2008})}\BibitemShut {NoStop}%
\end{thebibliography}%
\clearpage

\onecolumngrid
\beginsupplement
\setcounter{page}{1}

\section*{Supplemental Material}
\begin{center}
  \large
  \textbf{Measuring amount of computation done by C.elegans using whole brain neural activity} \\[1em]
  \normalsize
  Junang Li$^{1,*}$, Andrew M. Liefer$^{1,2}$, and David H. Wolpert$^{3,4,5,6,\dagger}$\\
  \textit{$^1$Department of Physics, Princeton University, Princeton, New Jersey 08544, United States of America}\\
  \textit{$^2$Department of Physics, Princeton University, Princeton, New Jersey 08544, United States of America}\\
  \textit{$^3$Santa Fe Institute, Santa Fe, New Mexico 87501, United States of America}\\
  \textit{$^4$International Center for Theoretical Physics, Trieste I-34151, Italy}\\
  \textit{$^5$Complexity Science Hub, Vienna 1080, Austria}\\
  \textit{$^6$Arizona State University, Tempe, Arizona 85287, United States of America}\\[1em]

  $^*$Corresponding author: \href{mailto:junangl@princeton.edu}{junangl@princeton.edu}
  $^\dagger$Corresponding author: \href{mailto:david.h.wolpert@gmail.com}{david.h.wolpert@gmail.com}
\end{center}

\begin{figure*}[h!]
    \centering
    \includegraphics[width=1\linewidth]{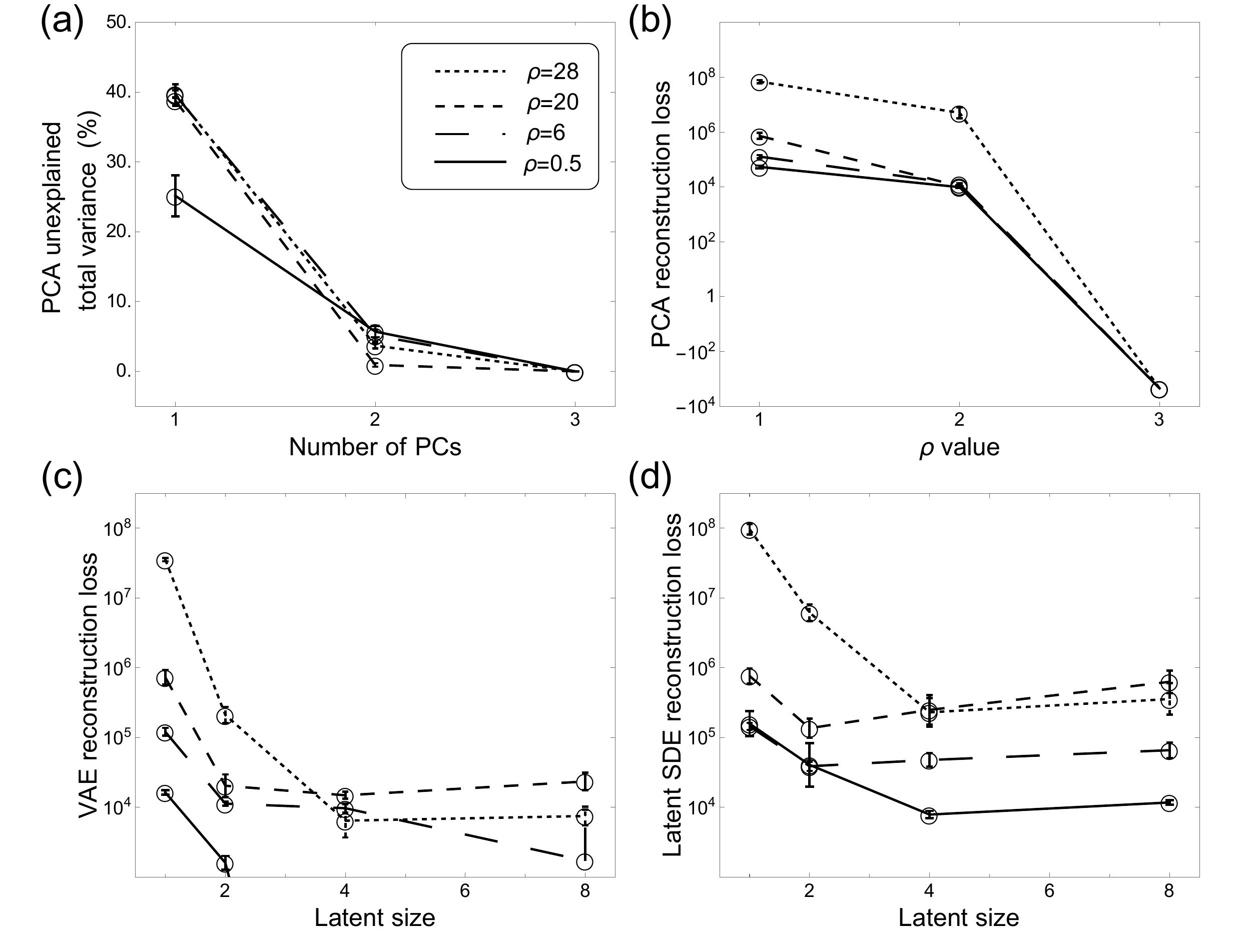}
    \caption{
    \textbf{Unexplained total variance and reconstruction loss with varying latent dimensions.}
    Since Lorenz dynamics is intrinsically three-dimensional, both unexplained total variance (a) and reconstruction loss (b) diminish at 3 principal components. However, for 1 and 2 principal components, the trend remains consistent.
    (c) Latent SDE reconstruction loss follows the same trend for 1 and 2 latent dimensions, with minimal differences observed between strange attractor and limit cycle behaviors when using 3 latent dimensions. 
    }
    \label{fig:lorenz_vary}
\end{figure*}

\begin{figure*}[h!]
    \centering
    \includegraphics[width=1\linewidth]{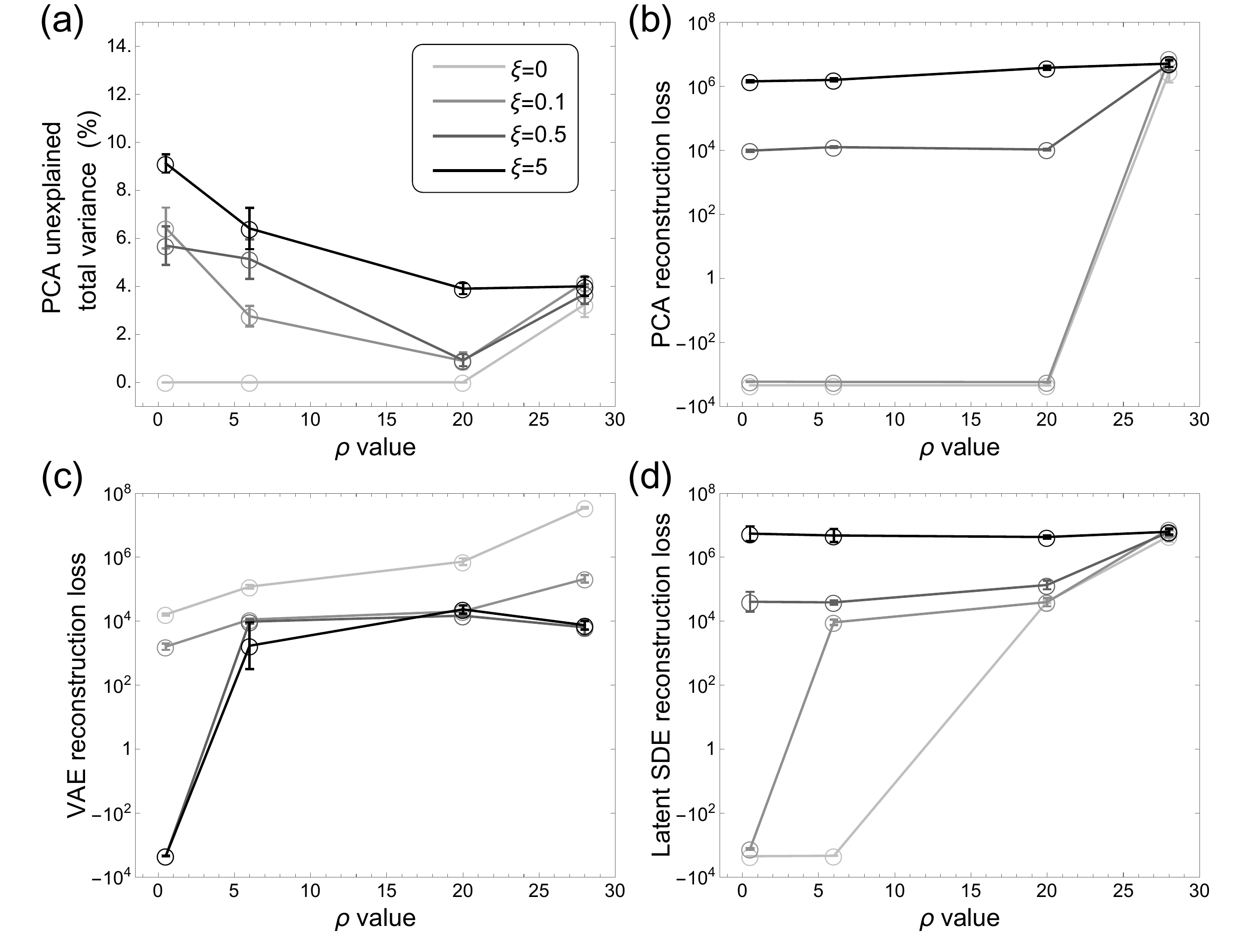}
    \caption{
    \textbf{Effect of noise level on the computation estimator.}
     Increasing noise levels generally lead to higher unexplained total variance and reconstruction loss. However, the impact is more pronounced for dynamics within stable regimes compared to chaotic strange attractors. Moreover, reconstruction loss kept the trend across different dynamical regimes with increased noise level. 
     (a) Total variance cannot be explained by the first 2 PCs. (b) Reconstruction loss when using the first 2 PCs.
     (c) Reconstruction loss when using first-order vector AR model. 
     (d) Reconstruction loss when using Latent SDE with 2 latent dimensions.  
    }
    \label{fig:lorenz_noise}
\end{figure*}

\begin{figure*}[h!]
    \centering
    \includegraphics[width=1\linewidth]{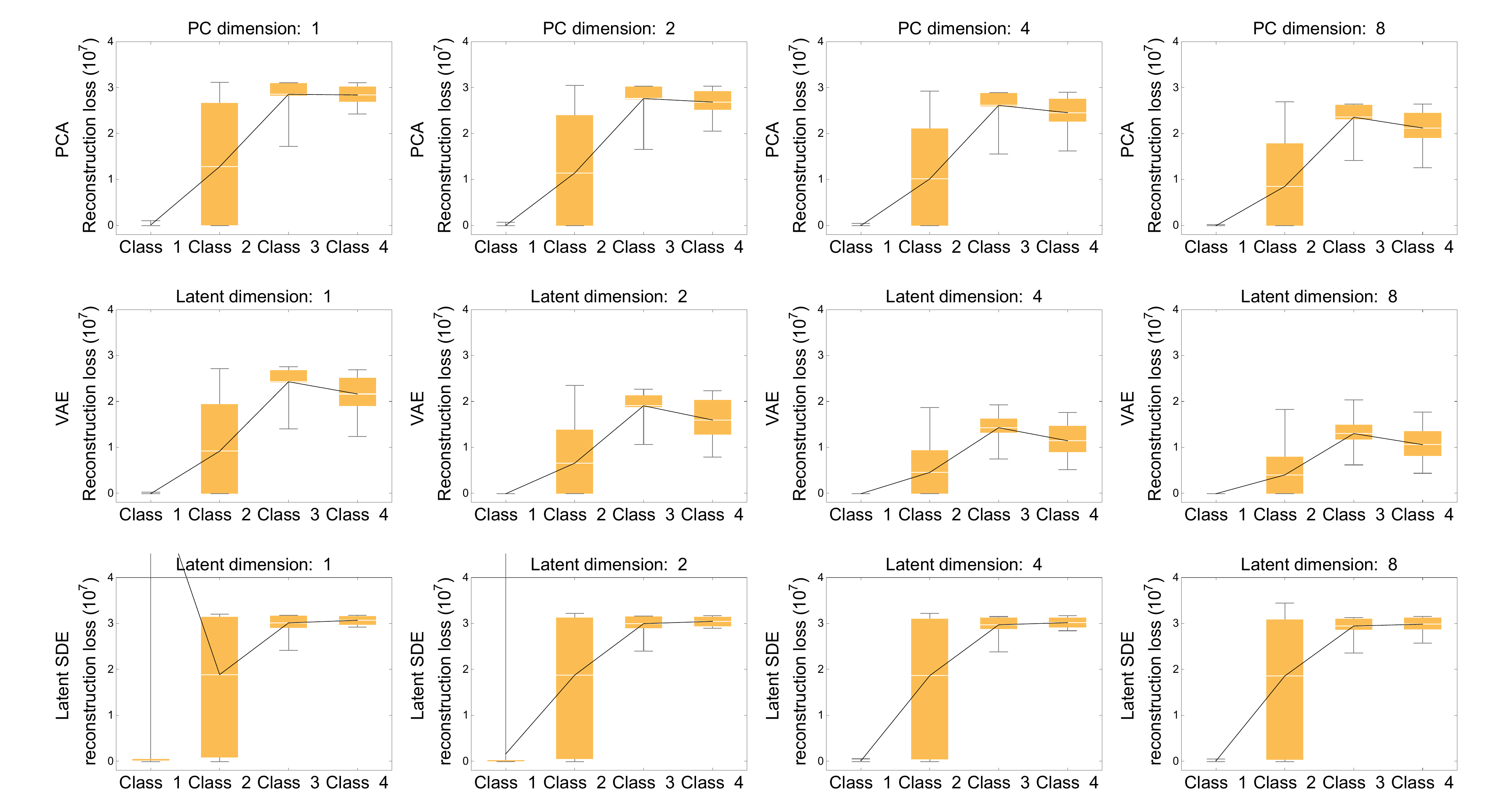}
    \caption{
    \textbf{Varying latent dimension does not change the trend of relative amount of computation measured different CA computation classes.}
    Top: Unexplained total variance.
    Middle: PCA reconstruction loss.
    Bottom: LatentSDE reconstruction loss. 
    Error bars indicate the maximum and minimum values, with the yellow bar representing the 25th to 75th percentile range. The mean is shown by the white line. Results represent all CA rules within the corresponding computation class, with each rule simulated across 10 independent random initial conditions
    }
    \label{fig:CA_vary}
\end{figure*}

\begin{figure*}[h!]
    \centering
    \includegraphics[width=1\linewidth]{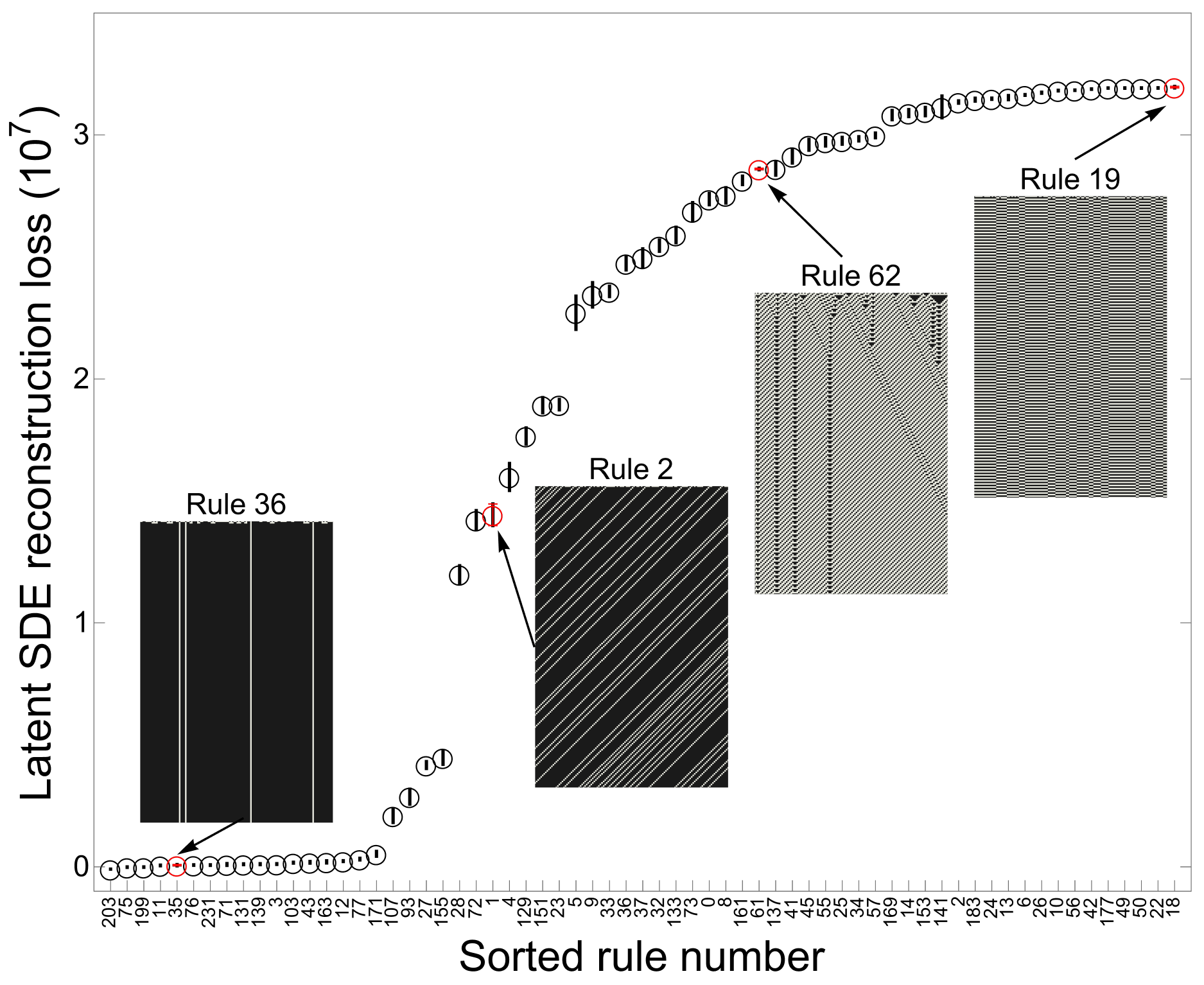}
    \caption{
    \textbf{The relative amount of computation within CA Class 2.}
    Using our method, we could distinguish the relative amount of computation within a CA computation class. 
    Sixty-five Class 2 rules are further categorized into subcategories, with examples of their dynamics shown as insets. 
    Generally, an increase in pattern complexity corresponds to an increase in reconstruction loss. 
    Notably, due to the continuous nature of the LatentSDE model, patterns exhibiting alternating oscillation dynamics yield the highest reconstruction losses. 
    Means and error bars indicate averages and standard errors over ten independent initial conditions. 
    }
    \label{fig:CA_class2}
\end{figure*}

\begin{figure*}[h!]
    \centering
    \includegraphics[width=1\linewidth]{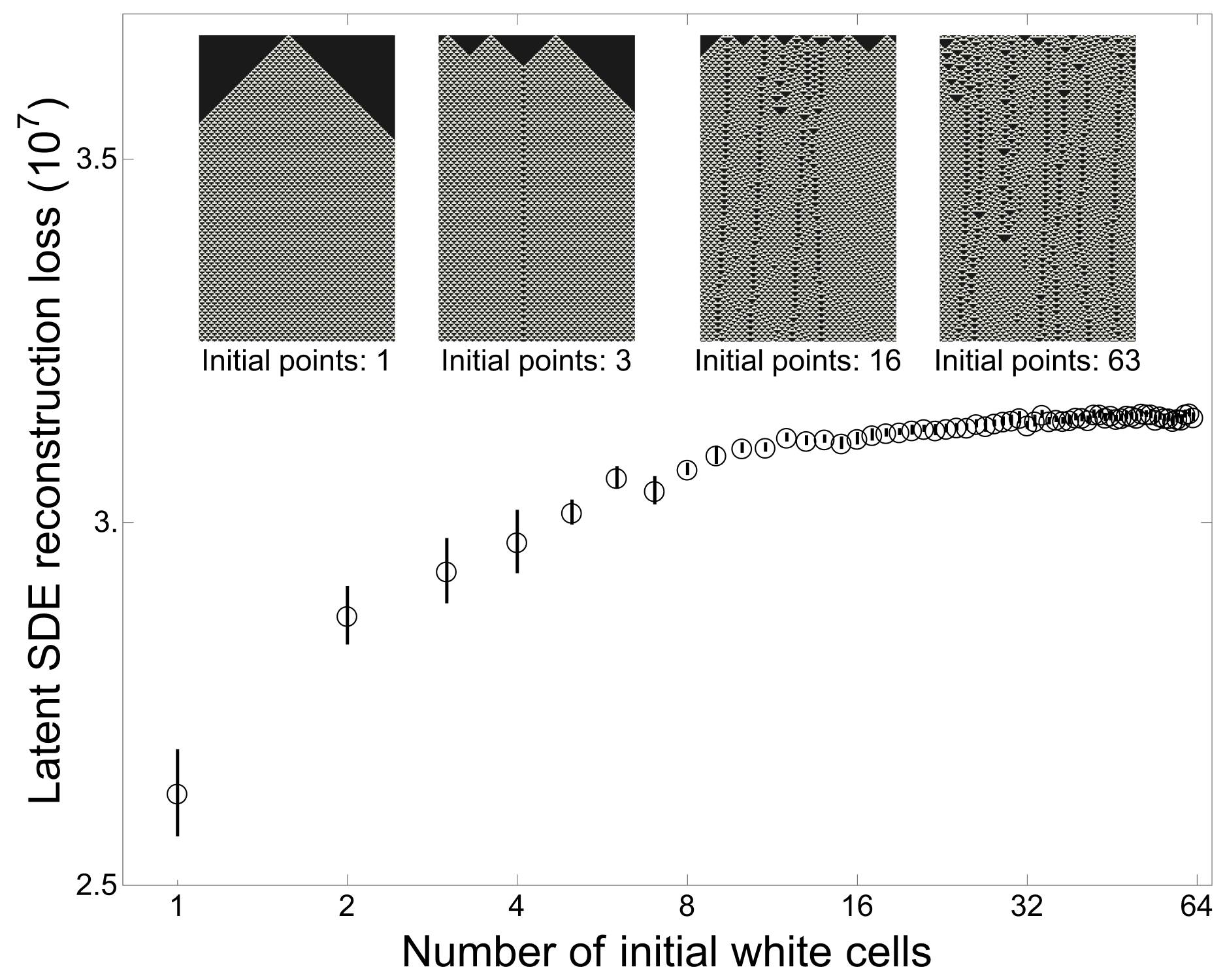}
    \caption{
    \textbf{The relative amount of computation for Rule 54 under different initial conditions.}
    Although Rule 54 is classified as Class 4, its dynamics do not necessarily exhibit Class 4 behavior under all initial conditions. 
    As shown in the insets, Rule 54 dynamics initialized with a small number of white cells resemble Class 2 dynamics. 
    Our method detects this dependency in the relative amount of computation, showing lower amount of computation for initial conditions with few white cells (effective Class 2 dynamics) and higher amount of computation for initial conditions with more white cells, where the dynamics more accurately reflect true Class 4 behavior.
    Means and error bars indicate averages and standard errors over ten independent initial conditions.
    }
    \label{fig:CA_rule54}
\end{figure*}

\begin{figure*}[h!]
    \centering
    \includegraphics[width=1\linewidth]{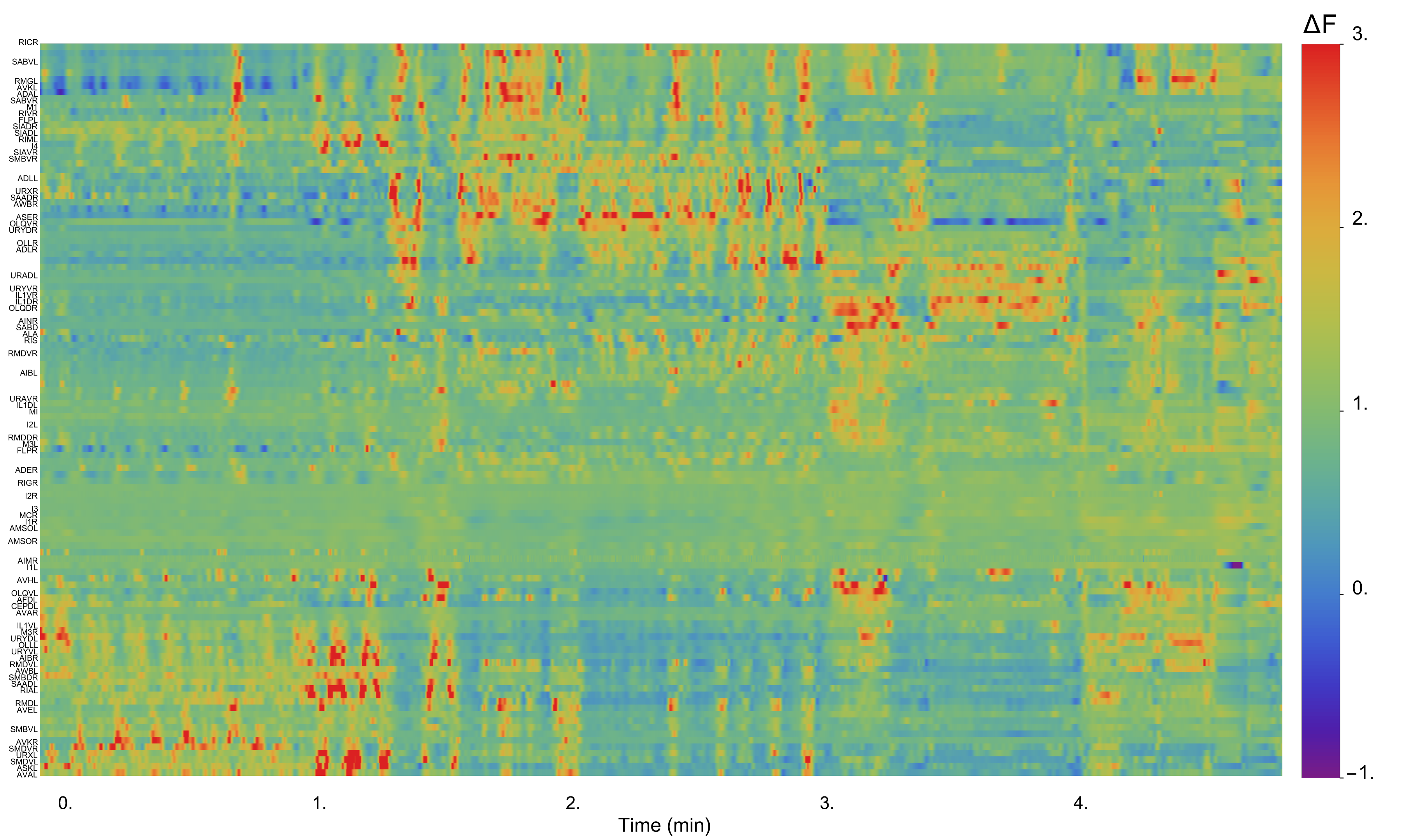}
    \caption{
    \textbf{NeuroPAL-identified whole-brain calcium imaging data.} 
    Neuron IDs are shown along the y-axis; missing labels indicate neurons that were not successfully identified or could not be matched with the calcium imaging data.
    }
    \label{fig:NeuroPAL}
\end{figure*}

\begin{figure*}[h!]
    \centering
    \includegraphics[width=1\linewidth]{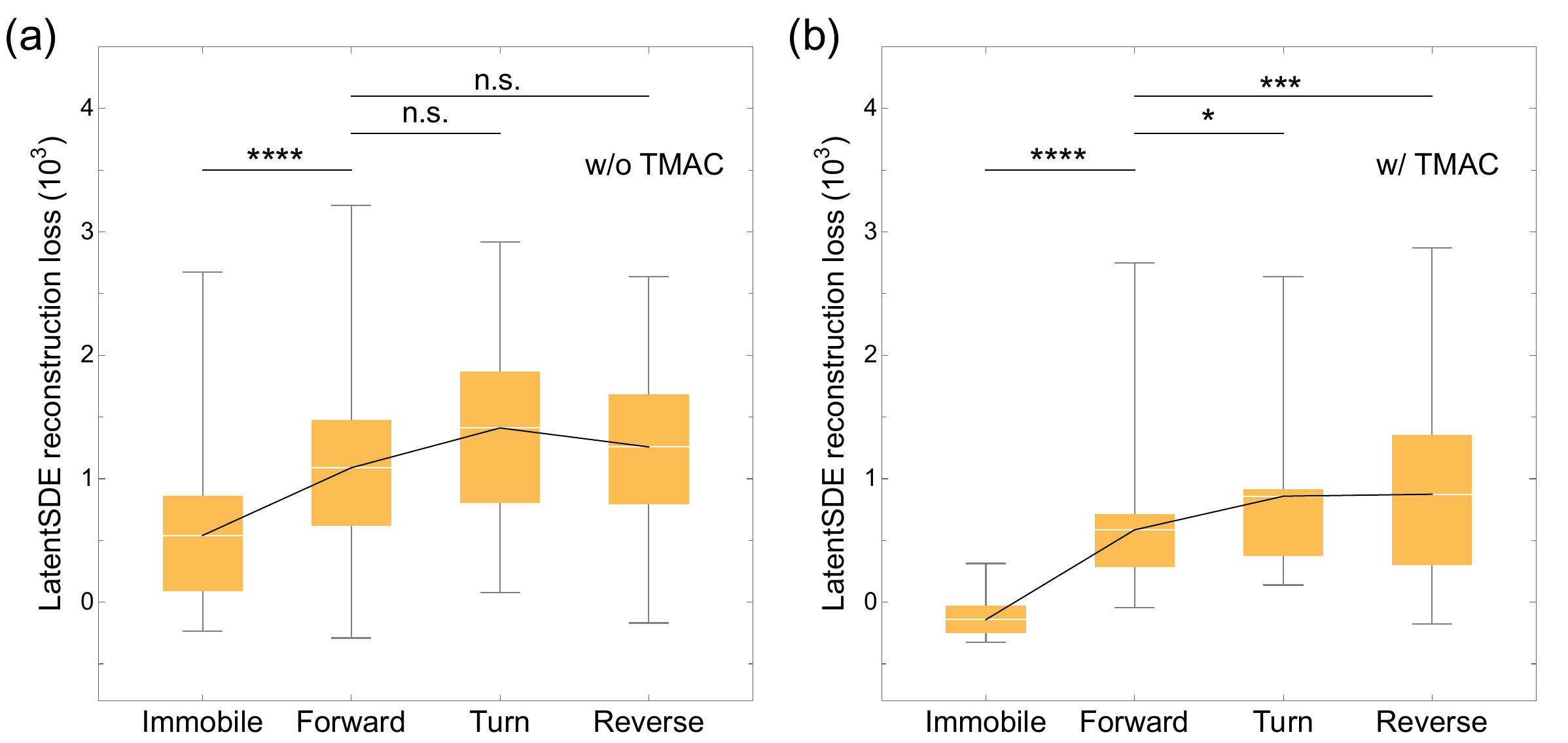}
    \caption{
    \textbf{Post analysis without TMAC cannot distinguish differences in locomotion states.}
    (a) Latent SDE reconstruction loss across different locomotion states from ratio metric without applying TMAC. The relative amount of computation associated with distinct locomotion states are indistinguishable. 
    (b) Latent SDE reconstruction loss across different locomotion states from TMAC, identical to the plot shown in Fig.~\ref{fig:worm_behavior}(d). 
    Error bars indicate the maximum and minimum values, with the yellow bar representing the 25th to 75th percentile range. The mean is shown by the white line.
    Statistical significance of the differences between behavioral states was assessed using t-tests, with significance levels indicated as follows: * for $p < 0.05$, ** for $p < 0.005$, *** for $p < 0.0005$, and **** for $p < 0.00005$.
    }
    \label{fig:ethogram_tmac}
\end{figure*}

\begin{figure*}[h!]
    \centering
    \includegraphics[width=1\linewidth]{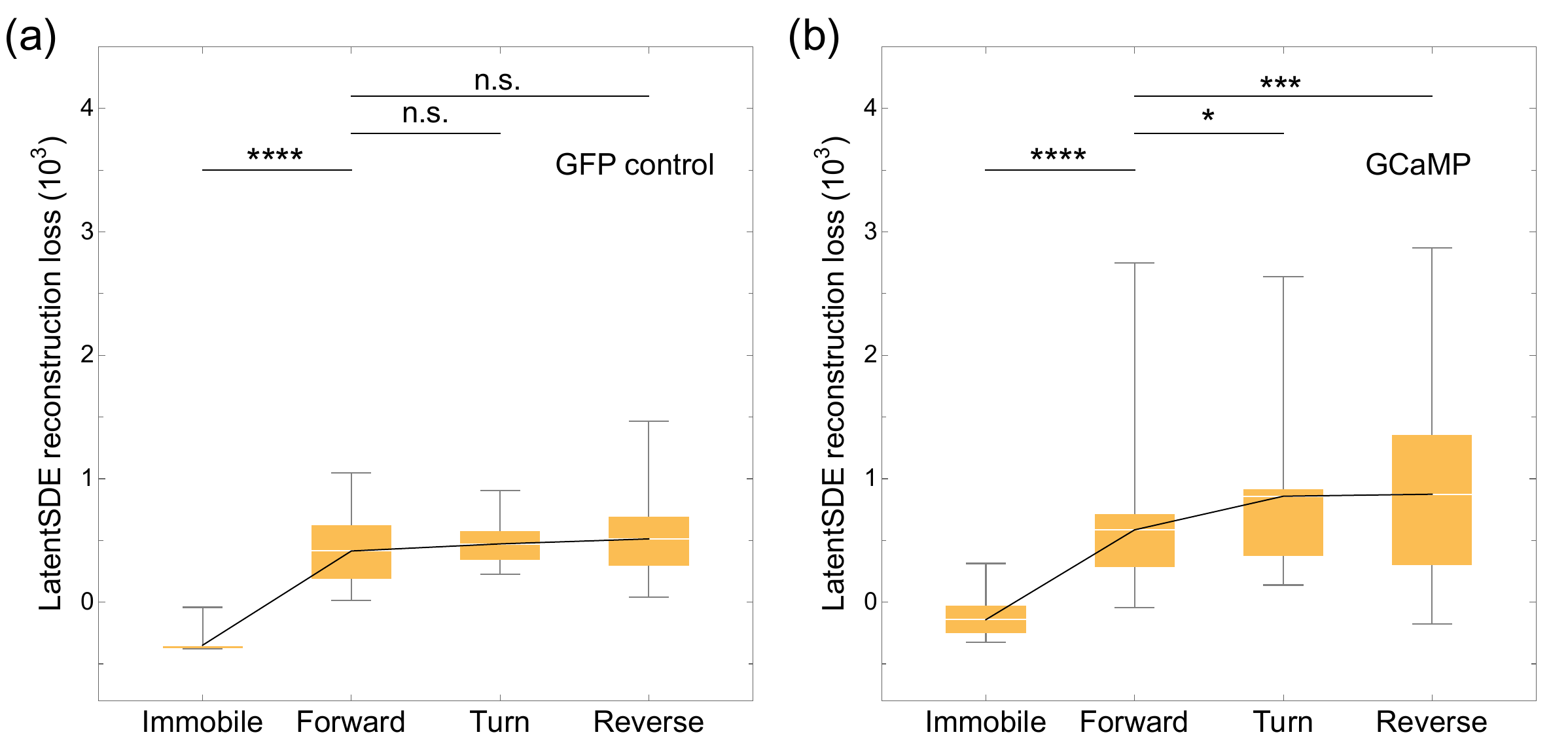}
    \caption{
    \textbf{GFP control experiments cannot distinguish differences in locomotion states.}
    Control experiments using calcium-insensitive GFP reveal the measurement noise in our experimental setup. 
    As expected, mobile worms exhibit higher noise levels, resulting in larger Latent SDE reconstruction losses. 
    However, this noise is consistent across different locomotion states and does not reflect the relative differences in the amount of computation associated with distinct behavioral states.
    (a) Latent SDE reconstruction loss across different locomotion states from GFP control signals.
    (b) Latent SDE reconstruction loss across different locomotion states from GCaMP signals, identical to the plot shown in Fig.~\ref{fig:worm_behavior}(d). 
    Error bars indicate the maximum and minimum values, with the yellow bar representing the 25th to 75th percentile range. The mean is shown by the white line.
    Statistical significance of the differences between behavioral states was assessed using t-tests, with significance levels indicated as follows: * for $p < 0.05$, ** for $p < 0.005$, *** for $p < 0.0005$, and **** for $p < 0.00005$.
    }
    \label{fig:ethogram_GFP}
\end{figure*}

\begin{figure*}[h!]
    \centering
    \includegraphics[width=1\linewidth]{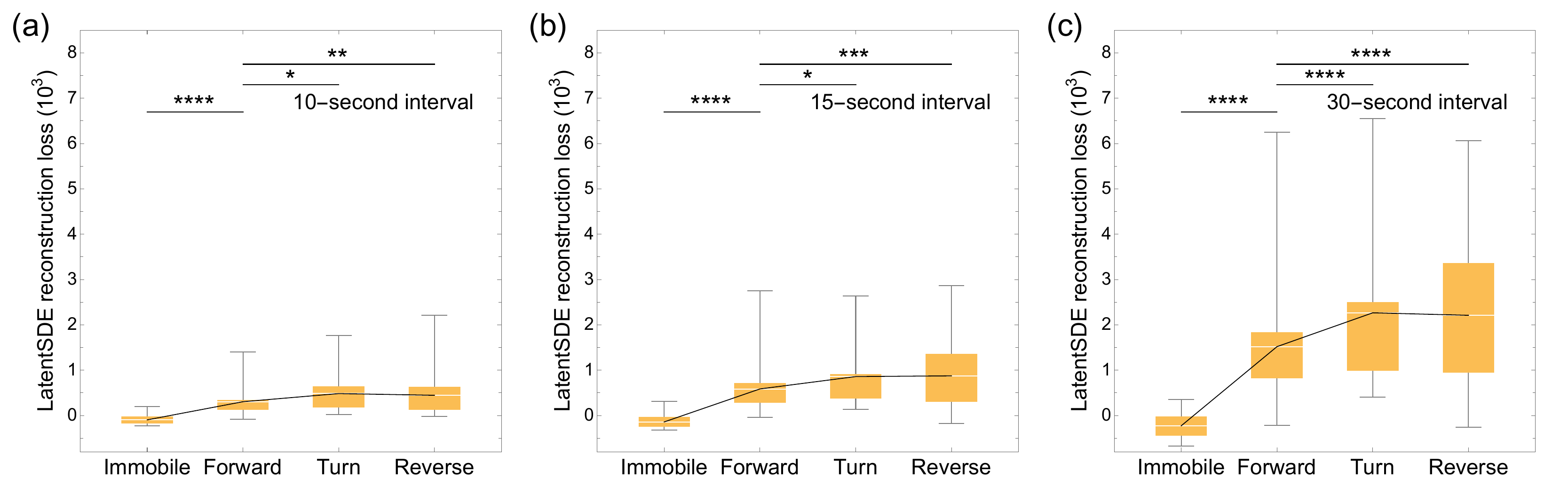}
    \caption{
    \textbf{Latent SDE reconstruction loss is robust to neural activity subsection length}
    (a) Latent SDE reconstruction loss across different locomotion states for neural recordings segmented into 10-second intervals.
    (b) Latent SDE reconstruction loss across different locomotion states for neural recordings segmented into 15-second intervals, identical to the plot shown in Fig.~\ref{fig:worm_behavior}(d).
    (c) Latent SDE reconstruction loss across different locomotion states for neural recordings segmented into 30-second intervals.
    Error bars indicate the maximum and minimum values, with the yellow bar representing the 25th to 75th percentile range. The mean is shown by the white line.
    Statistical significance of the differences between behavioral states was assessed using t-tests, with significance levels indicated as follows: * for $p < 0.05$, ** for $p < 0.005$, *** for $p < 0.0005$, and **** for $p < 0.00005$.
    }
    \label{fig:ethogram_datalen}
\end{figure*}

\begin{figure*}[h!]
    \centering
    \includegraphics[width=1\linewidth]{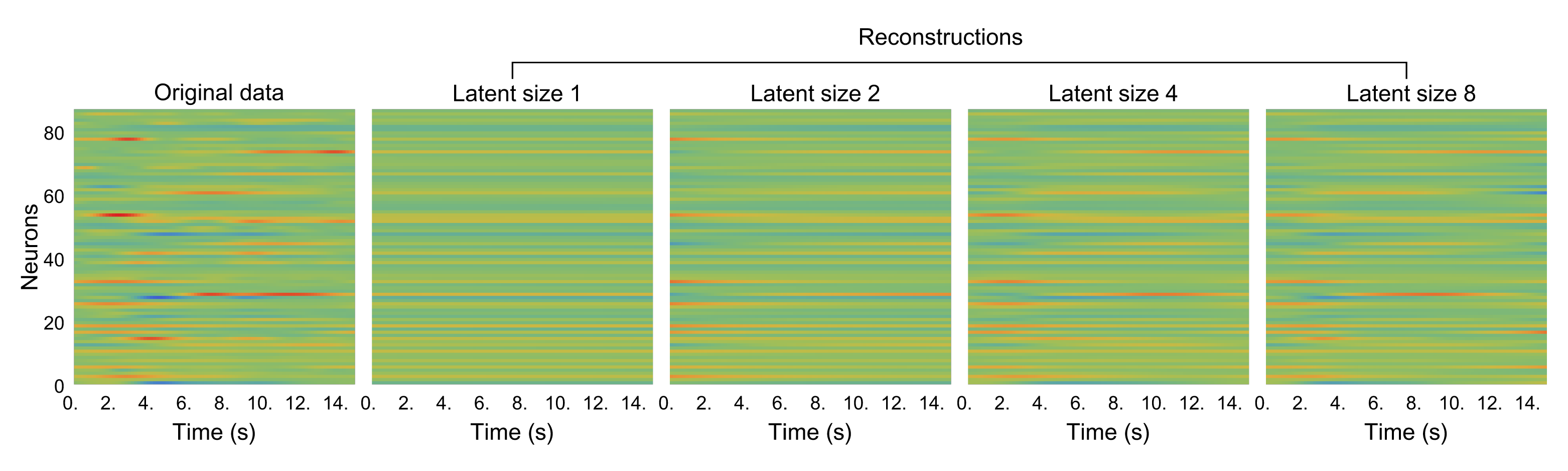}
    \caption{
    \textbf{Latent SDE reconstructions of neural activity.}
    The leftmost panel shows original neural activity pattern over a 15-second subsection. 
    The right panels display Latent SDE reconstructions of the same activity, illustrating the effects of increasing latent dimensionality.
    }
    \label{fig:reconstruction}
\end{figure*}

Additional figures and tables provide further validation and robustness checks: 
robustness of latent dimension (Figs.~\ref{fig:lorenz_vary}, and~\ref{fig:CA_vary}); 
effect of noise level on estimated computation (Fig.~\ref{fig:lorenz_noise}); 
detailed ranking of computation within CA Class 2 (Fig.~\ref{fig:CA_class2}); 
computation estimates for CA Rule 54 under varying initial conditions  (Fig.~\ref{fig:CA_rule54}); 
whole-brain imaging with NeuroPAL identification (Fig.~\ref{fig:NeuroPAL}); 
comparison without motion-correction algorithm (Fig.~\ref{fig:ethogram_tmac});
comparison with GFP control experiments (Fig.~\ref{fig:ethogram_GFP}); 
robustness of neural activity subsection length (Fig.~\ref{fig:ethogram_datalen}); 
and Latent SDE reconstruction of the neural activity (Fig.~\ref{fig:reconstruction}). 
Table~\ref{tab:methods} and~\ref{tab:worm} summarize reconstruction algorithm performance and \textit{C. elegans} strain details, respectively.

\clearpage

\clearpage

\clearpage

\clearpage

\clearpage

\clearpage

\clearpage

\clearpage

\clearpage

\clearpage

\clearpage

\clearpage
\begin{table}[h!]
    \centering
    \renewcommand{\arraystretch}{1.5} 
    \begin{tabular}{|>{\centering\arraybackslash}m{4cm}|>{\centering\arraybackslash}m{3cm}|>{\centering\arraybackslash}m{3cm}|>
    {\centering\arraybackslash}m{3cm}|>{\centering\arraybackslash}m{3cm}|}
        \hline
         & \textbf{PCA} & \textbf{VAE} & \textbf{VAR(1)} & \textbf{Latent SDE} \\ \hline
        \textbf{Stochastic Lorenz} & $\checkmark$ & $\checkmark$ & $\checkmark$ & $\checkmark$ \\ \hline
        \textbf{Cellular Automata} & $\times$ & $\times$ & $\times$ & $\checkmark$ \\ \hline
    \end{tabular}
    \caption{Performance comparison of different reconstruction algorithms (PCA, VAE, VAR(1), and Latent SDE) for quantifying the amount of computation across stochastic Lorenz dynamics and cellular automata..}
    \label{tab:methods}
\end{table}
\clearpage

\begin{table}[h!]
    \centering
    \renewcommand{\arraystretch}{1.5} 
    \begin{tabular}{|>{\centering\arraybackslash}m{4cm}|>{\centering\arraybackslash}m{4cm}|>{\centering\arraybackslash}m{4cm}|>{\centering\arraybackslash}m{4cm}|}
        \hline
        \textbf{Strain} & \textbf{Genotype} & \textbf{Role} & \textbf{Reference} \\
        \hline
        SFW 702 & flvIs17; otIs670 [low-brightness NeuroPAL]; lite-1(ce314); gur-3(ok2245) epochs & Calcium imaging & \cite{atanas2023brain} \\
        \hline
        AML 607 & wtfIs3[rab-3P::NLS::GFP::unc-54; rab-3P::NLS::tagRFP::unc-54]; otIs669[NeuroPAL] & GFP control & this work \\
        \hline
    \end{tabular}
    \caption{Strains used in this study}
    \label{tab:worm}
\end{table}

\end{document}